\documentclass[aps,prb,twocolumn,superscriptaddress,raggedbottom,showpacs,floatfix]{revtex4}   
\usepackage{amsmath,amsfonts,amssymb,amsthm}
\usepackage{color}
\usepackage{graphicx,latexsym}
\pdfoutput=1

\begin{document}
\title{Coulomb interaction and transient charging of excited states in open nanosystems}

\author{Valeriu Moldoveanu}
\affiliation{National Institute of Materials Physics, P.O. Box MG-7,
Bucharest-Magurele, Romania}
\author{Andrei Manolescu}
\affiliation{Reykjavik University, School of Science and Engineering, Kringlan 1,
IS-103 Reykjavik, Iceland}
\author{Chi-Shung Tang}
\affiliation{Department of Mechanical Engineering,  National United University,
Lienda, Miaoli 36003, Taiwan}
\author{Vidar Gudmundsson}
\affiliation{Science Institute, University of Iceland, Dunhaga 3, IS-107 Reykjavik,
Iceland}

\begin{abstract}

We obtain and analyze the effect of electron-electron Coulomb interaction
on the time dependent current flowing through a mesoscopic system
connected to biased semi-infinite leads.  We assume the contact is
gradually switched on in time and we calculate the time dependent reduced
density operator of the sample using the generalized master equation.
The many-electron states (MES) of the isolated sample are derived with
the exact diagonalization method.  The chemical potentials of the two
leads create a bias window which determines which MES are relevant
to the charging and discharging of the sample and to the currents,
during the transient or steady states.  We discuss the contribution
of the MES with fixed number of electrons $N$ and we find that in the
transient regime there are excited states more active than the ground
state even for $N=1$.  This is a dynamical signature of the Coulomb
blockade phenomenon.  We discuss numerical results for three sample
models: short 1D chain, 2D lattice, and 2D parabolic quantum wire.

\end{abstract}

\pacs{73.23.Hk, 85.35.Ds, 85.35.Be, 73.21.La}

\maketitle

\section{Introduction}

Due to the increasing interest in ultra-fast electron dynamics
considerable progress occurred recently in the theoretical
description of time dependent mesoscopic transport. New methods and
numerical implementations are rapidly evolving.  Transient currents
in open nanostructures are studied with Green-Keldysh formalism,\
\cite{Stefanucci,Moldo,Xiao} scattering theory,\cite{Gudmundsson} and
quantum master equation.\cite{Harbola,Welack1,NJP1,NJP2}  Most of the
results were obtained for noninteracting electrons due to the well known
computational difficulties to include time-dependent Coulomb effects.

It is nevertheless clear that the electron-electron interaction is
important in such problems.  An effort to incorporate it has been recently
done by Kurth {\it et al.}\cite{Kurth} followed by My\"oh\"anen {\it
et al.}\cite{Myohanen} who have described correlated time-dependent
transport in a short 1D chain defined by a lattice Hamiltonian.
The 1D sample was connected to external leads and the current was
driven by a time-dependent bias.  Those authors used a method based
on the Kadanoff-Baym equation for the non-equilibrium Green's function
combined with the time-dependent density functional theory to include
the Coulomb interaction in the sample.  Once the Green's functions were
calculated {\it total} average quantities of interest could be obtained,
like charge density or current, both in the transitory and in the steady
state. However this method does not say much about the dynamics of
specific {\it internal states} of the sample system.  In view of
the spectroscopy of excited states\cite{Tarucha} it is important to
have a theoretical tool for understanding separately the charging and
relaxation of the ground states and excited states in mesoscopic systems
in time-dependent conditions.

Our alternative is to use the statistical, or density operator.
The complete information about the time evolution of each quantum state
of the sample is captured in the reduced density operator (RDO), which
is the solution of the generalized master equation (GME).  Once the RDO
is defined in the Fock space the inclusion of the Coulomb interaction
becomes a known computational problem: obtaining the many-electron states
(MES) of the sample.  The RDO matrix is then calculated in the
basis of the interacting MES.

Let us enumerate some of the previous theoretical schemes to treat
transport and electron-electron interaction with the master equation. One
of the first attempts to derive a master equation for an interacting
system with time-dependent perturbations belongs to Langreth and
Nordlander for the Anderson model.\cite{Langreth} Gurvitz and Prager
started from the time-dependent Shr\"{o}dinger equation for the MES wave
functions and ended up with Bloch-like rate equations for the density
matrix of a quantum dot.\cite{Gurvitz} The electronic currents were calculated in the
steady state and it was shown that the Coulomb interaction renormalizes
the tunneling rates between the leads and the system.  In the same context
K\"{o}nig {\it et al.}\cite{real} developed a powerful diagrammatic
technique by expanding the RDO of a mesoscopic system in powers of the
tunneling Hamiltonian.  The time-dependence of the statistical operator
of the coupled and interacting system implies a quantum master equation
for the so called populations.  In this method the Coulomb interactions
are treated exactly, which makes it appealing for studying various
correlation effects like cotunneling.\cite{Becker}   The connection
between the real-time diagrammatic approach of  K\"{o}nig {\it et al.}
\cite{real} and the Nakajima-Zwanzig approach\cite{Nakajima,Zwanzig}
to the generalized master equation (GME) approach was made transparent
by Timm.\cite{Timm}

More recently Li and Yan\cite{Li} combined the $n$-resolved master
equation and the time dependent density-functional method to write down a
Kohn-Sham master equation for the reduced {\it single-particle} density
matrix.  Also, Esposito and Galperin,\cite{Esposito} using the equation
of motion for the Hubbard operators, have obtained a many-body description
of quantum transport in an open system and established a connection
between the GME and non-equilibrium Greeen's functions.  They studied simple
systems in the steady state regime: a resonant level coupled to a a single
vibration mode, an interacting dot with two spins, and a two-level bridge.
Another recent work by Darau {\it et al.}\cite{Darau} implemented the GME for a
benzene single-electron transistor and used exact MES to compute {\it
steady state} currents within the Markov approximation.\ \cite{Darau} The
stability diagram and the conductance peaks were obtained and a current
blocking due to interferences between degenerated orbitals was noticed.

In our previous papers\cite{NJP1,NJP2} we considered the GME method for
the RDO of independent electrons in the Fock space.  We discussed the
transient transport through quantum dots and quantum wires.  The contact
between the leads and the sample was switched on at a certain initial
moment $t_0$.  We discussed extensively the occupation of the states
within the bias window and the geometrical effects on the transient
currents.  We described the coupling between the sample and the leads
via a tunneling Hamiltonian in which we took into account the spatial
extension of the wave functions of both subsystems in the contact region.

In spite of earlier or more recent attempts a complete description of
the Coulomb effects in the time-dependent transport is still missing,
especially in sample models larger than a few sites.  In the present work
we combine the GME method with the Coulomb interaction in the sample and
we analyze the dynamics of the electrons starting with the moment when
the leads are coupled to the sample until a steady state is reached.
The Coulomb interaction is included in the Hamiltonian of the isolated
sample and the {\it interacting} MES are calculated with the exact
diagonalization method.  This means the Coulomb interaction is fully
included with no mean field assumption or density-functional model.
The number of single-electron states (SES) used to define the matrix
elements of the Hamiltonian of interacting electrons is sufficiently
large such that the MES of interest are convergent.  Due to the finite
bias window only a limited number of MES participate to the charge
transport through the sample, {\it i.\,e.}\ only those energetically
compatible with the electrons in the leads.  Hence the MES of interest
are selected by the chemical potentials in the leads.  We calculate
the RDO matrix elements in the subspace of these MES using the GME.
The electron-electron interaction in the leads is neglected.

It is well known that the Fock space increases exponentially with the
number of SES.  In addition the time dependent numerical solution of the
GME is also computational expensive. So at this stage we are limited to
describe only few electrons in the system: up to five in a small system,
but only up to three in a larger one.

The paper is organized as follows.  In Section 2 we briefly describe
the GME, the inclusion of the Coulomb interaction, and the selection of
the MES.  Next, in Section 3, we show results for three models: a short
1D chain, a 2D lattice of  12$\times$10 sites, and a finite quantum wire
with parabolic lateral confinement.  Conclusions and discussions are
presented in Section 4.

\section{ GME method and Coulomb interaction}

In this section we summarize the main lines of our method. The
equations apply both to the lattice and continuous models. The
time-dependent transport problem is considered within the partitioning
approach which is known both from the pioneering work of Caroli
\cite{Caroli} and from the derivation of the GME.  Prior to an initial
time $t_0$ the left lead (L) having a ``source'' role, and the right
lead (R) having a ``drain'' role, are not connected to the sample and
therefore can be characterized by equilibrium states with chemical
potentials $\mu_L$ and $\mu_R$ respectively.  Our aim is to compute
the time dependent currents flowing through the sample and leads
starting at moment $t_0$, when the three subsystems are connected,
until a stationary state is reached.

The generic Hamiltonian of the total system consisting of the sample
plus the leads is:
\begin{equation}\label{Htot}
H(t)=H_L+H_R+H_S+H_T(t) \,.
\end{equation}
$H_l$ with $l=L,R$ are the Hamiltonians of the leads. We denote by
$\varepsilon_{ql}$ and $\psi_{ql}$ the single-particle energies
and wave functions respectively, for each lead.  Using the creation
and annihilation operators associated to the single-particle states,
$c^{\dagger}_{ql}$ and $c_{ql}$, we can write
\begin{equation}\label{Hleads}
H_{l}=\int dq \, \varepsilon_{ql} \, c^{\dagger}_{ql}c_{ql} \,.
\end{equation}

$H_S$ is the Hamiltonian of the sample.  In the absence of the interaction the
SES have discrete energies denoted as $E_n$ and corresponding one-body
wave functions $\phi_n({\bf r})$.   Using now the creation and annihilation
operators for the sample SES, $d^{\dagger}_n$ and $d_n$, we can write
\begin{equation} \label{Hsample}
H_S=\sum_n E_n d^{\dagger}_nd_n + \frac{1}{2}\sum_{\substack{nm\\n'm'}}
V_{nm,n'm'}d^{\dagger}_nd^{\dagger}_md_{m'}d_{n'} \,.
\end{equation}
The second term in Eq.\ (\ref{Hsample}) is the Coulomb interaction.
In the SES basis the two-body matrix elements are given by:
\begin{equation}\label{Vcoul}
V_{nm,n'm'}=\int d{\bf r}d{\bf r'}\phi_n^*({\bf r})\phi_m^*({\bf r'})u({\bf r}-{\bf r'})
\phi_{n'}({\bf r})\phi_{m'}({\bf r'}),
\end{equation}
where $u({\bf r}-{\bf r'})$ is the Coulomb potential.

The third term of Eq.\ (\ref{Htot}) is the so-called tunneling Hamiltonian
describing the transfer of particles between the leads and the sample:
\begin{equation}\label{Htunnel}
H_T(t)=\sum_{l=L,R}\sum_n\int dq \chi_{l}(t)(T^l_{qn}c^{\dagger}_{ql}d_n+h.c.) \,.
\end{equation}
$H_T$ contains two important elements: (1) The time dependent switching
functions $\chi_l(t)$ which open the contact between the leads and the
sample; these functions mimic the presence of a time dependent potential
barrier. (2) The coupling $T^l_{qn}$ between a state with momentum $q$
of the lead $l$ and the state $n$ of the isolated sample, with wave
function $\phi_n$.  The coupling coefficients $T^l_{qn}$ depend on
the energies of the coupled states and, maybe more important, on the
amplitude of the wave functions in the contact region. As we have shown
in our previous work \cite{NJP1,NJP2} this construction allows us to
capture geometrical effects in the electronic transfer.  A precise
definition of the coupling coefficients is however model specific,
and will be mentioned in the next section.

The evolution of our system is completely determined by the statistical
operator $W(t)$ associated to the total Hamiltonian $H(t)$ defined in
Eq.(\ref{Htot}). $W(t)$ is the solution of the quantum Liouville
equation with a known initial value, prior to the coupling of
the sample and leads:
\begin{equation}\label{rho}
i \hbar \dot W(t)=[H(t),W(t)] \,, \quad W(t \leq t_0)=\rho_L \rho_R \rho_S \,,
\end{equation}
The isolated leads are described by equilibrium distributions,
\begin{equation}\label{rho_l}
\rho_l=\frac{e^{-\beta (H_l-\mu_l N_l)}}{{\rm Tr}_l \{e^{-\beta(H_l-\mu_l N_l)}\}}
\,, \quad l=L,R \,,
\end{equation}
and the isolated sample by the density operator $\rho_S$.  After the coupling
moment the dynamics of the sample is conveniently described by the
RDO which is defined by averaging the total statistical operator over
those degrees of freedom belonging to the leads:
\begin{equation}\label{redrho}
\rho(t)={\rm Tr}_L {\rm Tr}_R W(t),\quad \rho(t_0)=\rho_S.
\end{equation}

In the absence of the electron-electron interaction the MES
eigenvectors of $H_S$ are bit-strings of the form $| \nu \rangle =
|i_1^{\nu},i_2^{\nu},..,i_n^{\nu}...\rangle$, where $i_n^{\nu}=0,1$
is the occupation number of the $n$-th SES.  The set $\{\nu \}$ is a
basis in the Fock space of the isolated sample and the RDO can be seen
as a matrix in this basis.  From Eqs.\ (\ref{rho})-(\ref{redrho}) we
obtain in the lowest (2-nd) order in the coupling parameters $T^l_{qn}$
the GME (see Ref.\ \onlinecite{NJP1} for details):
\begin{eqnarray}\nonumber
{\dot\rho}(t)=&-&\frac{i}{\hbar}[H_S,\rho(t)]\\\label{GMEfin}
&-&\frac{1}{\hbar^2}\sum_{l=L,R}\int dq\:\chi_l(t)
([{\cal T}_{ql},\Omega_{ql}(t)]+h.c.) \,,
\end{eqnarray}
where the coupling operator ${\cal T}_{ql}$ has matrix elements
\begin{equation}\label{Tmunu}
\left( {\cal T}_{ql} \right)_{\mu\nu}=\sum_n T^l_{qn}\langle \mu | d^{\dagger} | \nu \rangle \,.
\end{equation}
The operators $\Omega_{ql}$ and $\Pi_{ql}$ are defined as
\begin{eqnarray}\nonumber
&&\Omega_{ql}(t)=e^{-itH_S} \int_{t_0}^tds\:\chi_l(s)\Pi_{ql}(s)e^{i(s-t)\varepsilon_{ql}}e^{itH_S},\\\nonumber
&&\Pi_{ql}(s)=e^{isH_S}\left ({\cal T}_{ql}^{\dagger}\rho(s)(1-f_l)-\rho(s){\cal T}_{ql}^{\dagger}f_l\right )e^{-isH_S}
\end{eqnarray}
and $f_l$ is the Fermi function of the lead $l$.

In the presence of the electron-electron interaction in the sample
the MES which are eigenstates of $H_S$ are linear combinations of
bit-strings: $H_S | \alpha ) = \cal E_{\alpha} | \alpha )$, where
$| \alpha ) = \sum_{\nu} C_{\alpha\nu} | \nu \rangle$, $C_{\alpha\nu}$
being the mixing coefficients which can be found together with the
energies $\cal E_{\alpha}$ by diagonalizing $H_S$.  (To distinguish
better between the noninteracting and the interacting MES we use the right
angular bracket for the former and the regular curved bracket for the
later.)  Using now the set $\{\alpha\}$ as a basis, {\it i.\,e.}\
the {\em interacting} MES, the GME has the same form as Eq.\ (\ref{GMEfin}),
where the matrix elements of all operators are now defined in the
interacting basis and the matrix elements of the coupling operators are
\begin{equation}\label{Tmunuint}
\left( {\cal T}_{ql}\right)_{\alpha\beta}=\sum_n T^l_{qn}(\alpha|d^{\dagger}|\beta) \,.
\end{equation}

Because the sample is open the number of electrons $N$ contained in the
sample is not fixed.  The Hamiltonian $H_S$ given in  Eq.\ (\ref{Hsample})
commutes with the total ``number'' operator $\sum_n d_n^{\dagger} d_n$.
Thus $N$ is a ``good quantum number''  such that any state $|\alpha )$ has
a fixed number of electrons.  So the MES can also be labeled as
$|\alpha )=|N,i)$ with $i=0,1,2,...$ an index for the ground and excited states
of the MES subset with $N$ electrons.  The many-body energies can also
be written as ${\cal E}_{\alpha}={\cal E}^{(i)}_{N}$. In the practical
calculations $N$ varies between 0 (the vacuum state) and $N_{max}$ which is
the total number of SES considered in the numerical diagonalization of $H_S$.
The total number of MES is thus $2^{N_{max}}$.

If the coupling between the leads and the sample is not too strong we
expect that only a limited number of MES participate effectively to the
electronic transport.  These states are naturally selected by the bias
window $[\mu_R, \mu_L]$.  In the following examples, by selecting suitable
values of the chemical potentials in the leads, we will truncate the basis
of interacting MES to a reasonably small subset such that we can solve
numerically Eq.\ (\ref{GMEfin}) with our available computing resources.
To relate the bias window with the effective MES we need to consider
the chemical potential of the isolated sample containing $N$ electrons,
\begin{equation}
\mu^{(i)}_N={\cal E}^{(i)}_N-{\cal E}^{(0)}_{N-1 \,,}
\end{equation}
which is the energy required to add the $N$-th electron on top
of the ground state with $N-1$ to obtain the $i$-th MES with $N$
particles.\cite{Vaz} We expect the current associated to the MES
$|N,i)$ to depend on the location of the chemical potential $\mu^{(i)}_N$
relatively to the bias window.  In particular it is clear that if at
the coupling moment $t_0$ the sample is empty all MES with $\mu^{(i)}_N \gg
\mu_L$ will remain empty both during the transient and the steady states,
so they can be safely ignored when solving the GME.

\section{ Models and results}

We have numerically implemented the GME method both for lattice and continuous models.
The sample models are: a short 1D chain with 5 sites, a 2D rectangular lattice with
$12\times 10=120$ sites, and a short quantum wire withe parabolic lateral confinement.
In all cases the coupling functions have the form
\begin{equation}\label{Chi}
\chi_l(t)=1-\frac{2}{e^{\gamma t}+1}
\end{equation}
with $\gamma$ a constant parameter, such that at the initial moment,
which is $t_0=0$, we have $\chi_l(0)=0$ (no coupling), and in the steady
state, for $t\to\infty$, $\chi_l=1$ (full coupling).

\subsection{ A toy model: short 1D chain}

In this model the two semi-infinite leads are attached to the ends of a
1D chain with 5 sites.  The coupling between a lead state with wave function
$\psi_{ql}$ and a sample state with wave function $\phi_n$
is given by the product between the wave functions at the contact site:
\begin{equation}\label{Tqn}
T^{l}_{qn}=V_l{\psi}^{*}_{ql}(0)\phi_n(i_l) \,,
\end{equation}
where 0 is the contact site of the lead $l=L,R$, the end sites of the sample
being $i_L=1$ and $i_R=5$.
\begin{figure}[tbhp!]
\includegraphics[width=0.40\textwidth]{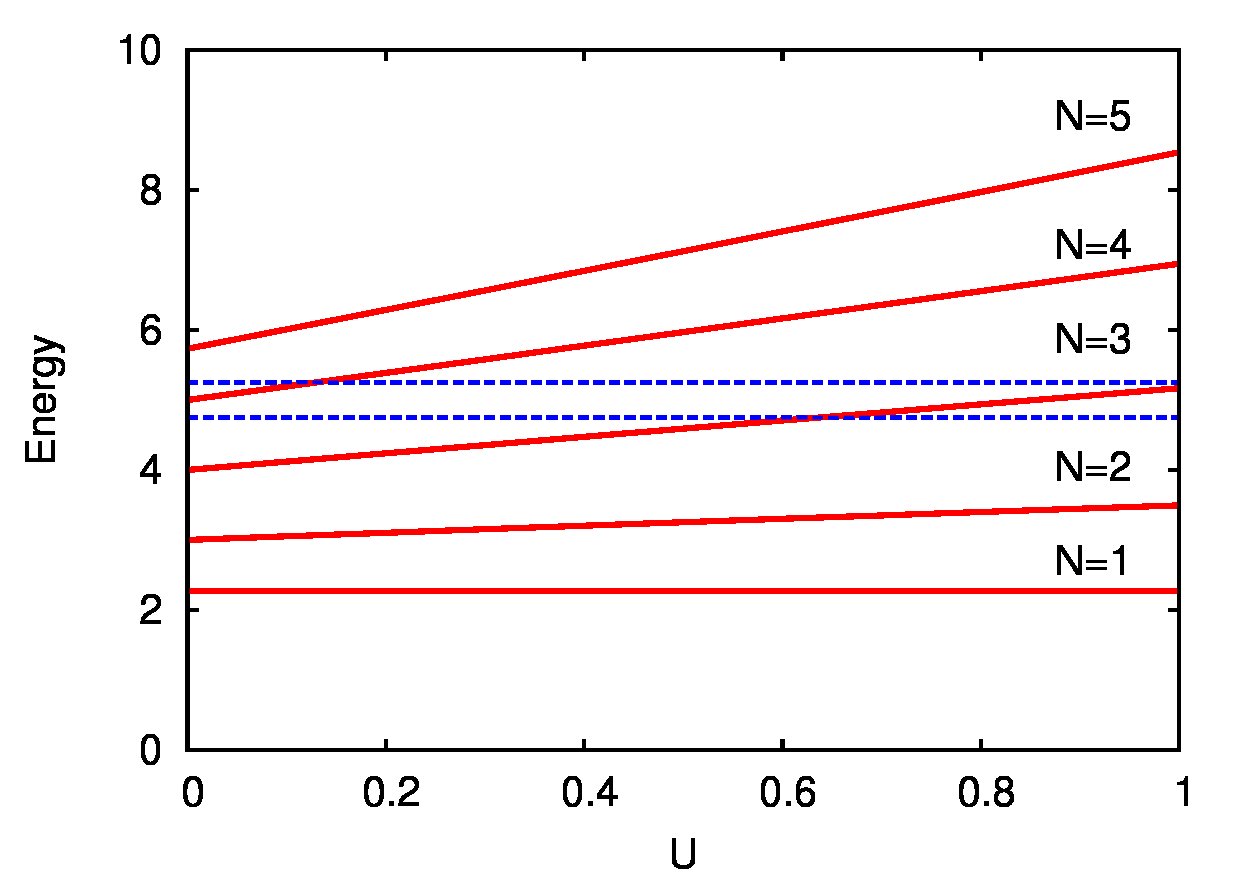}
\caption{(Color online) The equilibrium chemical potentials
$\mu^{(0)}_N$ for $1\leq N \leq 5$ as a function of the interaction strength $U$.
The dotted lines mark the chemical potentials of the leads selected in the transport
simulations shown in the next figure, i.\,e.\ $\mu_L=5.25$ and $\mu_R=4.75$. }
\label{figure1}
\end{figure}

The reason to call this a toy model is that we can obtain the complete
set of $2^5=32$ MES, {\it i.\,e.}\ we do not need to cut the basis of the
5 SES.  We also do not need to cut the MES basis, all matrix elements
of the statistical operator can be numerically calculated, even if not
all of them might be important for the currents.  In addition we will
consider the strength of the Coulomb interaction as a free parameter $U$,
whereas in a realistic systems this is fixed by the electron charge
and the dielectric constant of the material.  Our goal is to have a
qualitative understanding of the underlying physics, and in particular to
show the presence of the Coulomb blocking effects  at certain values
of $U$ or of the chemical potentials of the leads.  The Coulomb matrix
elements defined in Eq.\ (\ref{Vcoul}) are calculated as
\begin{equation}\label{Vcoul1D}
V_{nm,n'm'}=\sum_{i \neq i'}\phi_n^*(i)\phi_m^*(i')\,\frac{U}{\mid i-i'\mid}\,\phi_{n'}(i)\phi_{m'}(i')\,.
\end{equation}

In Fig.1 we show the equilibrium chemical potentials $\mu^{(0)}_N$
corresponding to ground states with $1\leq N \leq 5$ particles against the
interaction strength $U$. One observes a linear dependence of $\mu^{(0)}_N$
on $U$, with slope increasing with $N$.  Obviously the total Coulomb energy
increases both with $U$ and $N$.

Let us now briefly review the Coulomb blockade
scenario.\ \cite{Kouwenhowen} Suppose the isolated sample contains $N$
electrons and the chemical potentials of the leads are chosen such that
$\mu^{(0)}_{N}<\mu_R < \mu_L<\mu^{(0)}_{N+1}$. Then the bias window
$[\mu_R,\mu_L]$ may include one or more of the excited configurations
with $N$ particles.  In general some states with $N$ electrons may have
excitation energies exceeding $\mu_L$ or even $\mu^{(0)}_{N+1}$. This
situation corresponds to the Coulomb blockade phenomenon. Indeed,
the addition of the $(N+1)$-th electron is energetically forbidden.
Consequently the current in the steady state should vanish.  However,
shorter or longer transient currents are generated by all many-body
configurations in the vicinity of the bias window.
\begin{figure}[tbhp!]
\includegraphics[width=0.45\textwidth]{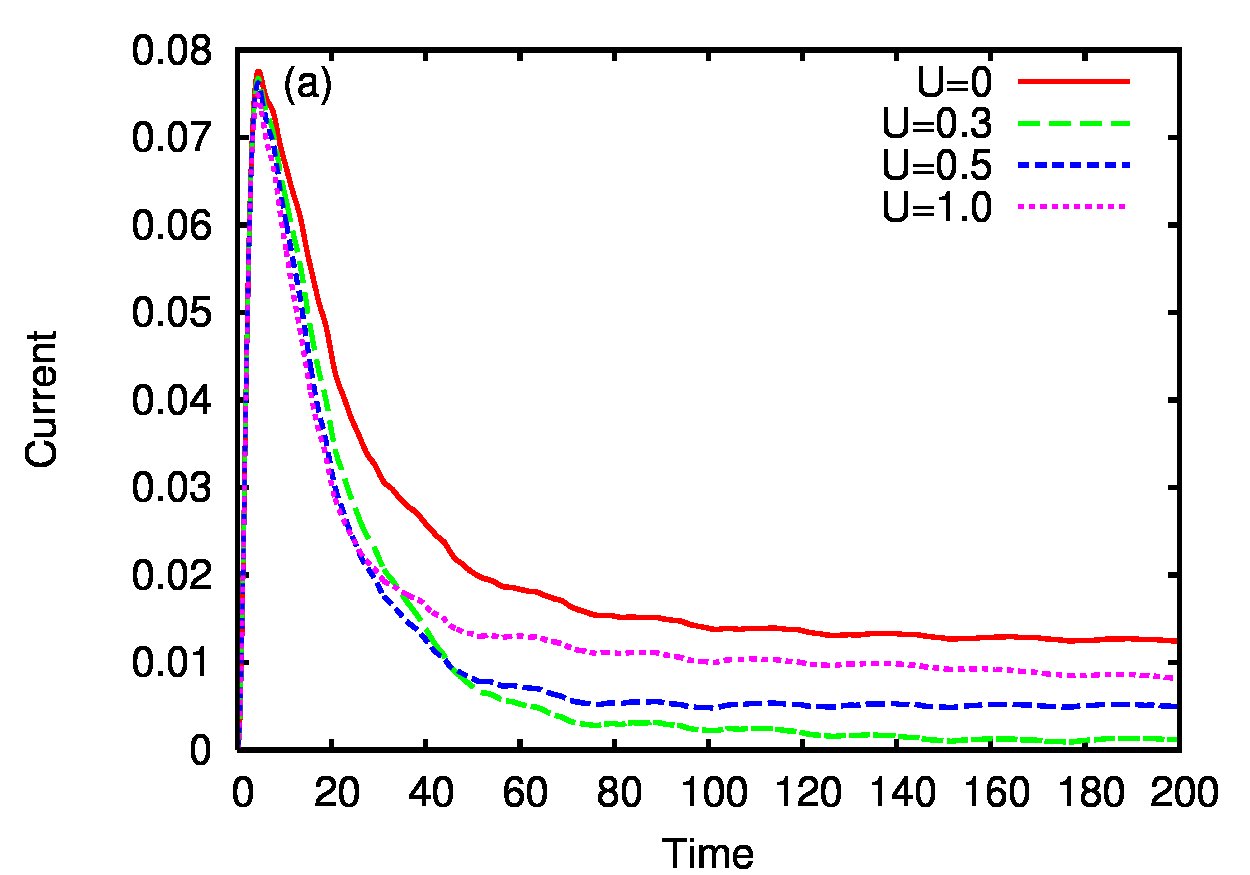}
\includegraphics[width=0.45\textwidth]{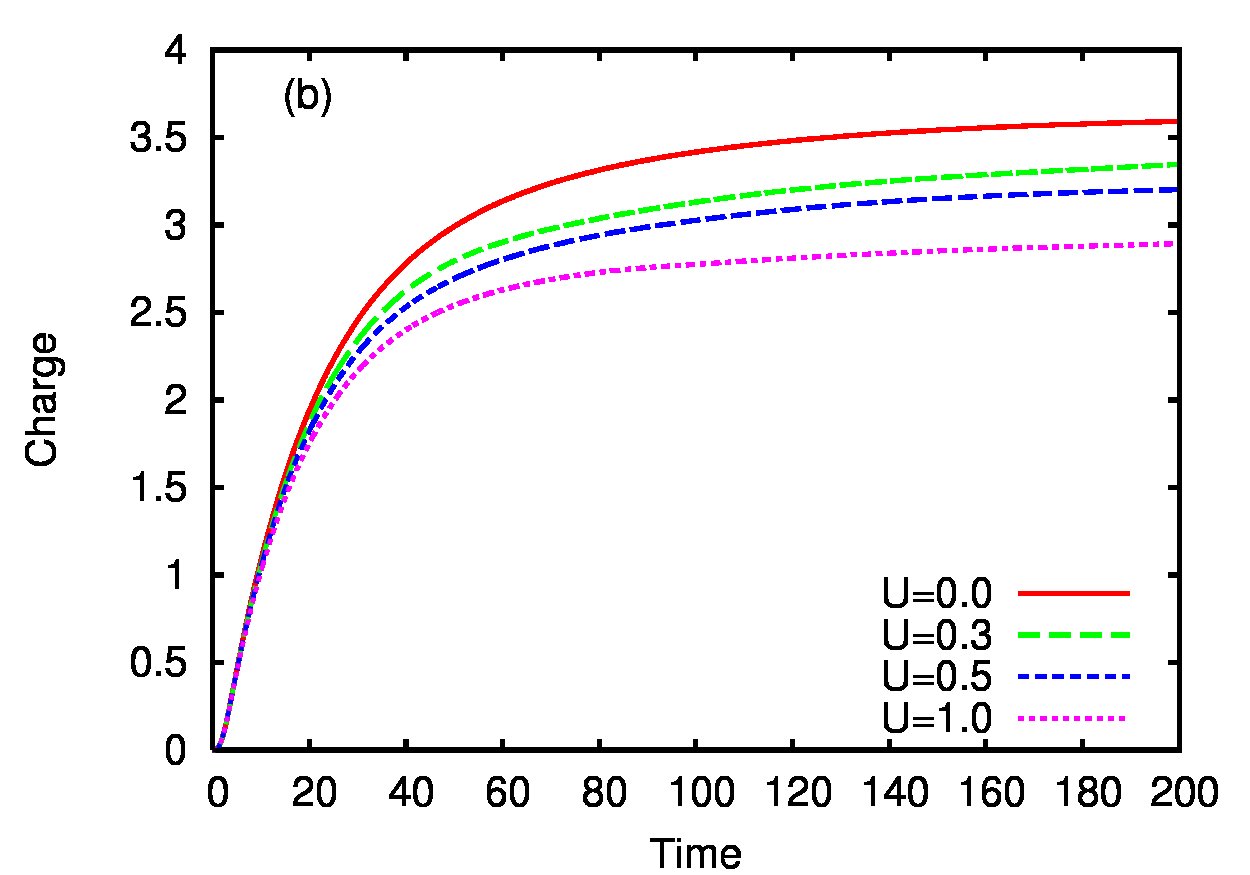}
\caption{(Color online) The total current entering the $5\times 1$ sample from the left lead
as a function of time for the different values of the interaction strength $U$.
The chemical potentials of the leads $\mu_L=5.25$ and $\mu_R=4.75$. }
\label{figure2}
\end{figure}

Fig.\ 2(a) and 2(b) show the total currents in the left lead and the
total charge residing in the sample for several values of the interaction
strength.  $U$ is measured in units of $t_S$, the hopping parameter in
the sample,\ \cite{NJP1} and the time is expressed in units of $\hbar/t_S$
while the current is in units of $et_S/\hbar$.  The coupling constant
in Eq.\ (\ref{Chi}) is $\gamma=1$.  The system is initially empty and
thus $\rho(0)=|00000\rangle\langle 00000|$.

The chemical potentials of the leads, $\mu_L=5.25$ and $\mu_R=4.75$,
are chosen such that in the absence of Coulomb interaction, {\it i.\,e.}\
for $U=0$, $\mu^{(0)}_4$ is located within the bias window. In this
case we obtain in the steady state the  mean number of electrons about
3.6 and a non-vanishing current in the leads.  This is understandable,
since $\mu^{(0)}_4=E_4=5$, which is the 4-th level of the isolated
sample. The occupation of this level in the steady state is about 0.6,
the other states being either full or empty.  Also in this case, the
excited states have small contributions to the steady state current
as the system tends to be in the ground state with $N=3$ electrons.
Those contributions may also depend on the coupling strength of individual
states with the leads, but in general remain small.\cite{Valim2}

The situation may change for $U\neq 0$. For the interacting system,
{\it e.\,g.} for $U=0.3$, the system settles down in the Coulomb
blockade regime, the total current being almost suppressed in the
steady state. This happens because the interaction pushes the chemical
potentials upwards such that for $U=0.3$ both ground states with $N=3$
and $N=4$ electrons are outside the bias window and cannot produce
steady currents. When the interaction strength is further increased to
$U=0.5$ and $U=1$ the steady state currents are gradually restored. This
could look surprising, but one can see in Fig.1 that by increasing $U$
the ground state configuration with 3 electrons approaches and enters
the bias window. Consequently the transport becomes again possible.
Note that while the steady state currents are not monotonous w.r.t. $U$
the charge absorbed in the system continuously decreases, Fig.\ 2(b).
\begin{figure}[tbhp!]
\includegraphics[width=0.45\textwidth]{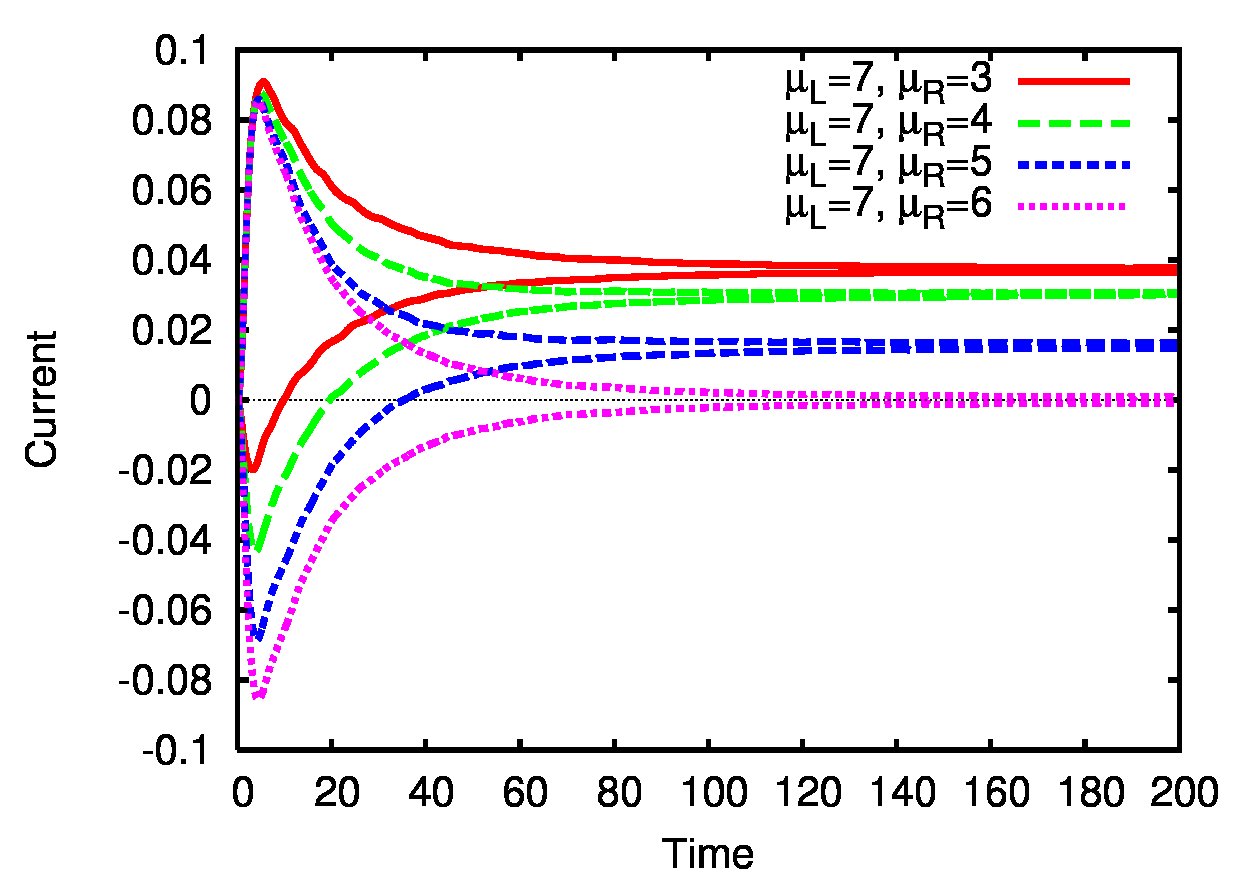}
\caption{The time-dependent total currents in the left and right leads at
different values of the chemical potential $\mu_R$. The current in the right
lead starts at negative values.  Other parameters: $V_L=V_R=0.750$, $U=1.0$.}
\label{figure3}
\end{figure}

In transport experiments the strength of the electron-electron interaction
is indeed fixed. The usual way to obtain the Coulomb blockade is to vary
the chemical potentials of the leads relatively to the energy levels of
the sample, or vice versa.  In Fig.\ 3 we show the currents in both
leads for different values of the chemical potential $\mu_R$, while
keeping fixed $\mu_L=7$. The strength of the Coulomb interaction is $U=1$
and $\mu^{(0)}_4$ almost equals $\mu_L$. The steady state value of the
current decreases as $\mu_R$ increases, because fewer states are included
in the bias window.  The Coulomb blockade onset occurs for $\mu_R>5$,
when $\mu^{(0)}_3$ drops below $\mu_R$. We observe that the maximum value
of the total current in the left lead does not change much when $\mu_R$
varies. In contrast, the transient current in the right lead is negative
and increases in magnitude as $\mu_R$ increases.  This means that the right
lead feeds the many-body configurations that fall below $\mu_R$.

The contribution of the excited states to the transient and steady
state currents depends strongly on the bias window.  In Fig.\ 4 we
show the currents entering the sample from the left lead, carried
by the states with $N=2$ and $N=3$ electrons, for $\mu_R=3,4,5$ (the
cases with non-vanishing current in the steady state).  We also show
separately the contribution to the currents associated to the ground
state configurations, related to $\mu^{(0)}_2$ and $\mu^{(0)}_3$, and
the complementary contribution of all the excited states with 2 and 3
particles.  In this case the wave vectors of the ground states are mostly
given by the non-interacting wave vectors: $|11000\rangle$ with weight
97\% and $|11100\rangle$ with 98\% for $N=2$ and $N=3$ respectively.
\begin{figure}[tbhp!]
\includegraphics[width=0.237\textwidth]{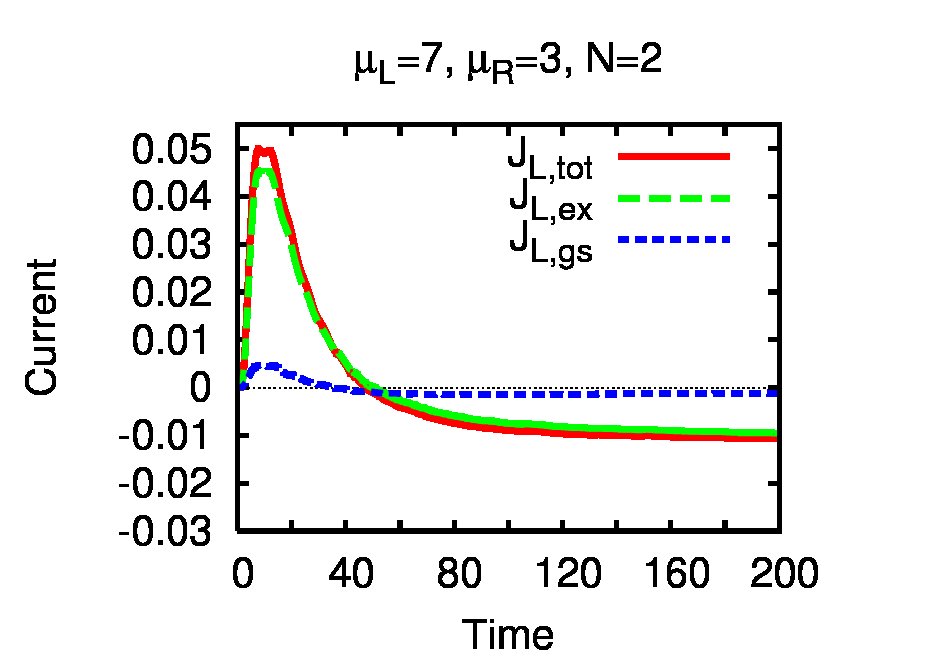}
\includegraphics[width=0.237\textwidth]{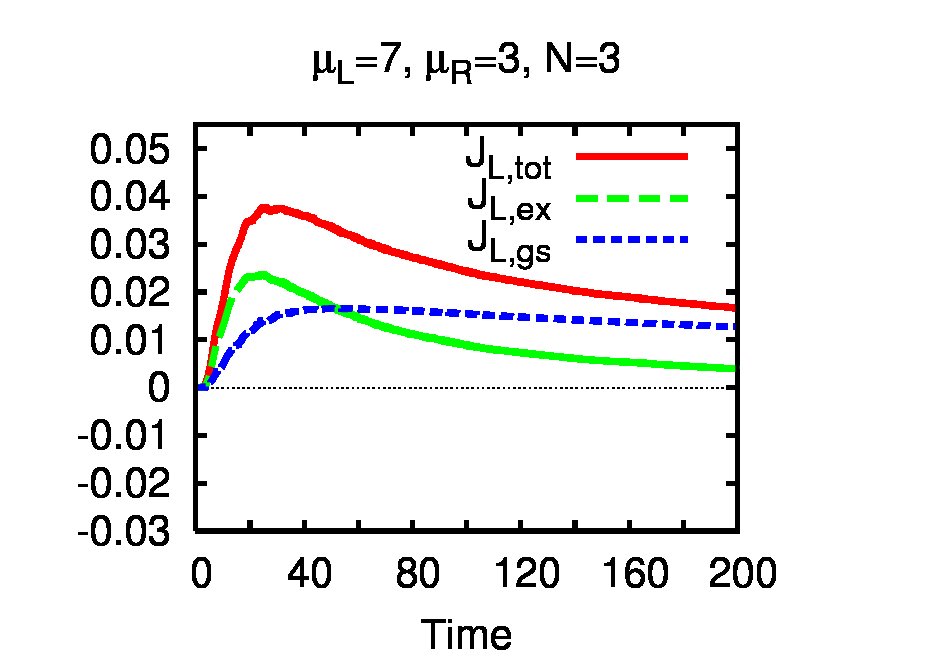}
\includegraphics[width=0.237\textwidth]{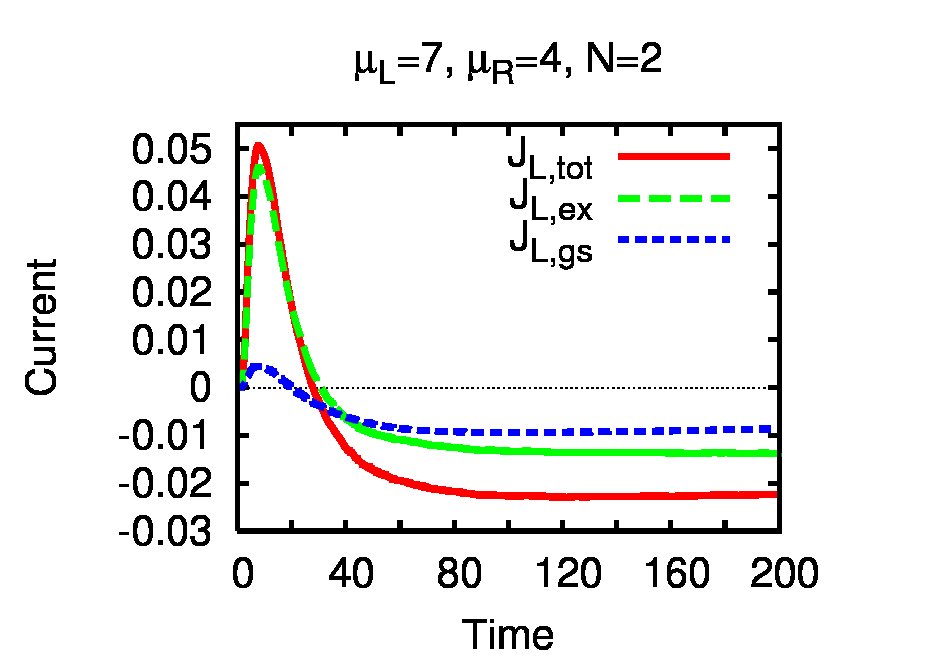}
\includegraphics[width=0.237\textwidth]{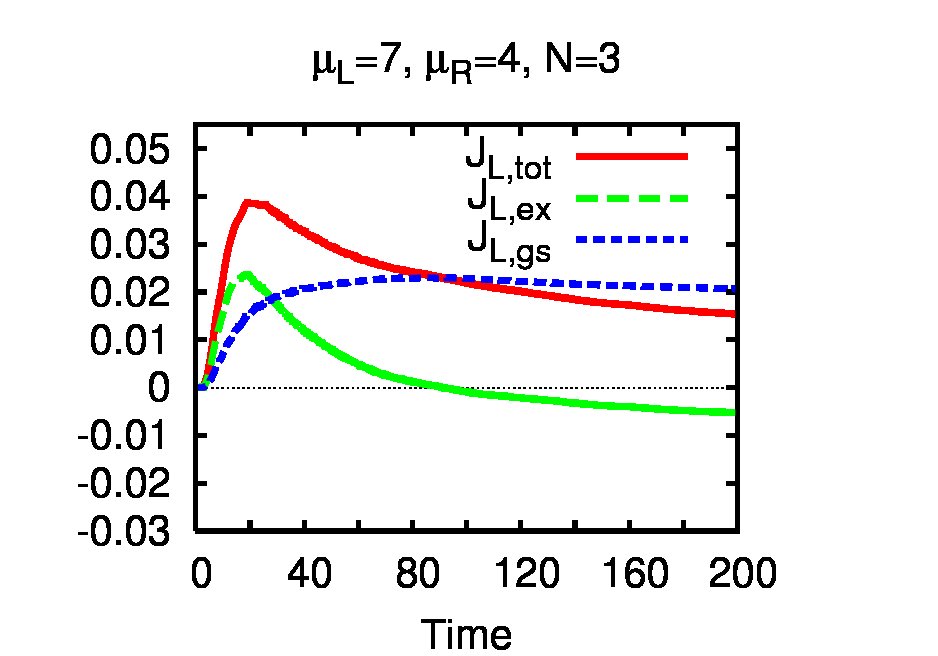}
\includegraphics[width=0.237\textwidth]{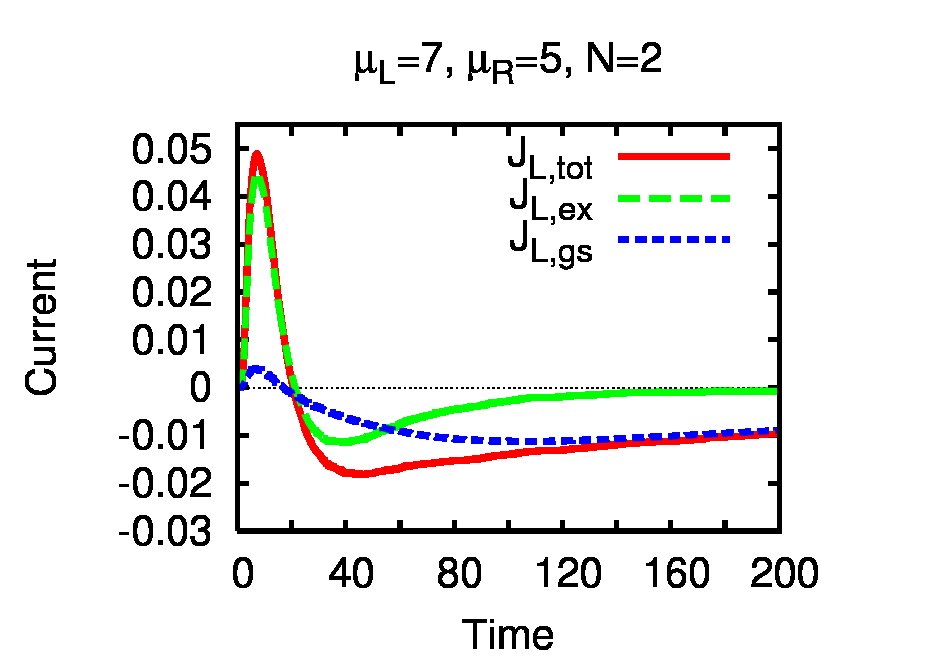}
\includegraphics[width=0.237\textwidth]{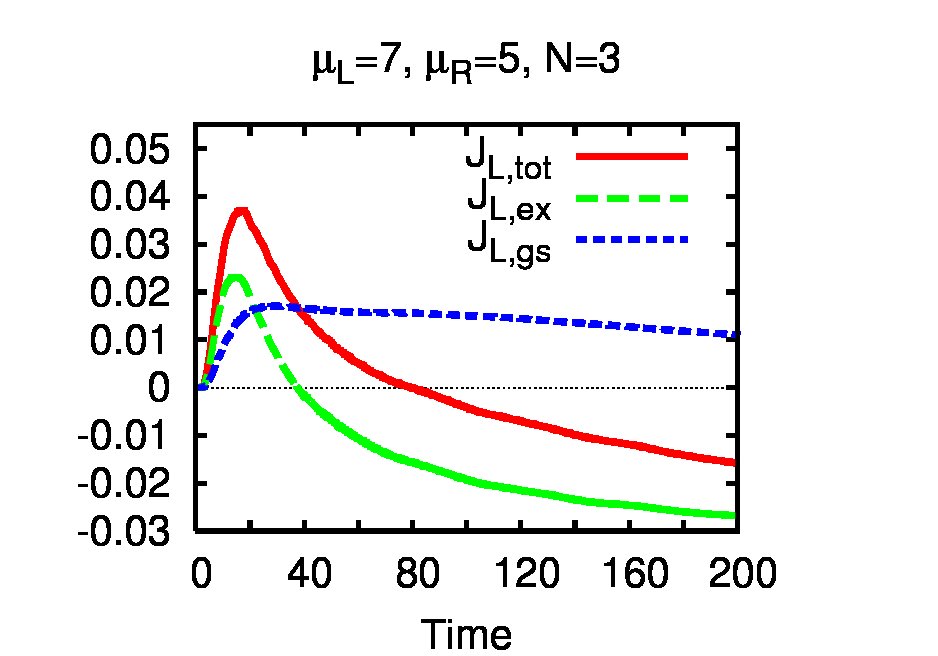}
\caption{The separate contributions to the current of the ground state with $N$
particles and of {\it all} excited states with $N$ particles, for different
values of $\mu_R$. For completeness we also include the total currents $J_L$
for the same configurations. The discussion is made in the text.
Other parameters: $V_L=V_R=0.750$, $U=1.0$. }
\label{figure4}
\end{figure}

For $\mu_R=3$ the steady state current of the ground state configuration
is vanishingly small and so the total negative current associated
to two-particle states comes mostly from the excited states. In the
many-body energy spectrum of the isolated sample we obtain 5 excited
configurations with $\mu^{(i)}_2 \in [\mu_R,\mu_L]=[3,7]$.  As $\mu_R$
moves up the steady state current of the ground state with $N=2$ becomes
also negative. The combined contributions of the excited states vanishes
at $\mu_R=5$. As can be seen from Fig.\ 1 $\mu_R=5$ is well above
$\mu^{(0)}_2$, but very close to $\mu^{(0)}_3$.  Consequently, the ground
configuration with $N=2$ is heavily populated in the steady state,
whereas the excited states have low probability and thus weak current.
Actually, as we have checked, all the currents associated to each excited
state with $N=2$ vanish individually.  In the transient regime however the
$N=2$ currents in all three cases are dominated by the excites states.

The currents of the excited states having $N=3$ electrons are positive
at $\mu_R=3$, but change sign at $\mu_R=4$. For $\mu_R=5$ their magnitude
exceeds the contribution of the ground state which is always positive.
A more detailed analysis of the currents carried by specific excited
states will be given for the 2D model.

Finally, both in the transient and in the steady states the currents have
small periodic fluctuations determined by the permanent transitions
of electrons between the states in the sample and the states in the
leads and back.\cite{Valim2} They are best seen in Fig.\ 2(a).
Such fluctuations have also been obtained very recently by Kurth {\it
et al.} using combination of the non-equilibrium Green's functions
and the time dependent density-functional theory of the Coulomb
interaction.\cite{Kurth2}

\subsection{2D lattice}

We show now results for a 2D rectangular lattice with $12\times10$
sites. For a lattice constant of $a=5$ nm this sample can be seen
as a discrete version of a quantum dot of 60 nm $\times$ 50 nm. We
used the lowest 10 SES of the non-interacting sample in the numerical
diagonalization of the interacting Hamiltonian. This number is sufficient
to produce convergent results for the first 50 MES for an interaction
strength $U=0.8$.  The Coulomb matrix elements are calculated in the
same way as for the 1D case, Eq.\ (\ref{Vcoul1D}), except that now
the site indices are two-dimensional, {\it i.\,e.} $i=(i_x,i_y)$ and
$i^\prime=(i^{\prime}_x,i^{\prime}_y)$.

The two contact sites are chosen at diagonally opposite corners of the
sample.  The coupling coefficients are calculated with Eq.\ (\ref{Tqn}),
like for the 1D chain, and depend on the wave function of the particular
SES at the contact sites. These coefficients are illustrated in
Fig.\ 5(a). The reduced density matrix is calculated using the first
50 MES. This allowed us to take into account many-body configurations
with up to 3 electrons.

In Fig.\ 5(b) we show the chemical potentials $\mu^{(i)}_N$ for the
ground and excited states with $N=1,2,\ \mathrm{and}\ 3$ particles. At
the initial moment $t_0=0$ the system is empty.  Based on the previous
example, the main contribution to the currents in the steady states is
expected from those MES with ground state chemical potentials
located inside the bias window $[\mu_R,\mu_L]$.  One also observes
excited configurations with $N$ particles having chemical potentials
larger than $\mu^{(0)}_{N+1}$.

\begin{figure}[tbhp!]
\includegraphics[width=0.45\textwidth]{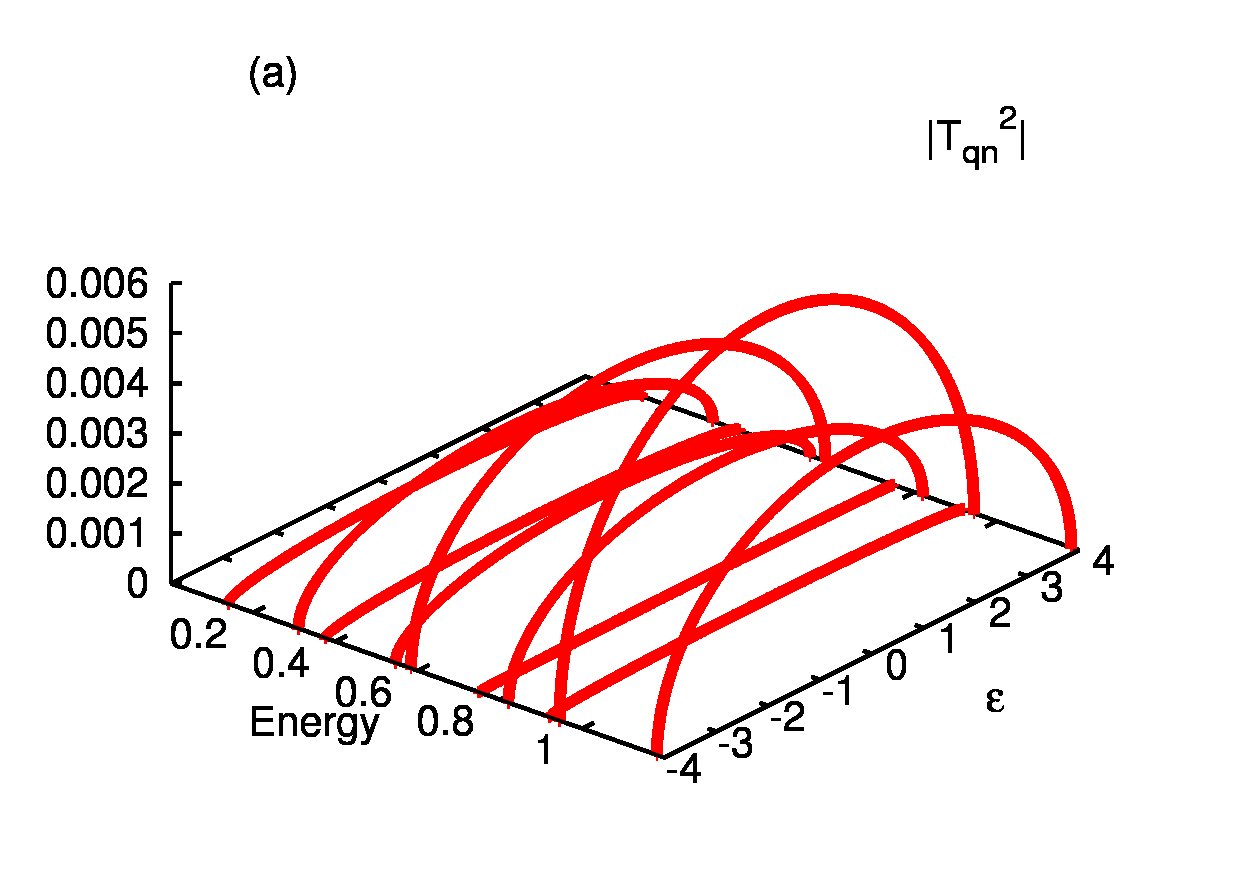}
\includegraphics[width=0.45\textwidth]{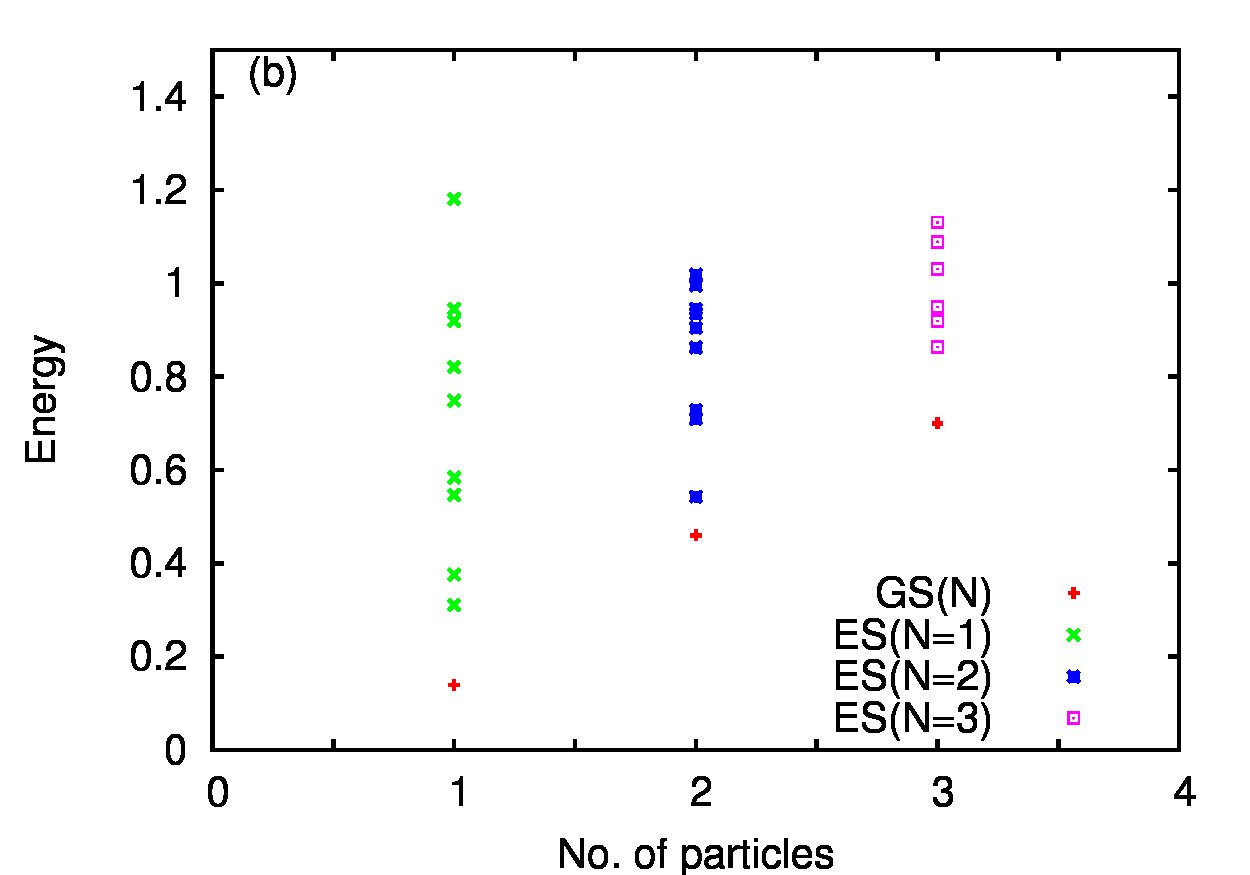}
\caption{(Color online) (a) The coupling amplitudes $|T_{qn}|^2$ for $n=1,..,5$ between single-particle
states in the leads with momentum $q$ and the lowest 5 single-particle states of the isolated dot.
(b) The generalized chemical potentials for $N$ -particle interacting configurations. The red crosses
mark  $\mu^{(0)}_N$ while the other ones correspond to generalized potentials
$\mu^{(i)}_N$ related to the $i$-th excited state of the $N$ particle system.}
\label{figure5}
\end{figure}

In the following we discuss the currents carried by the various many-body
states involved in transport.  In a first series of calculations we
selected the chemical potential $\mu_R=0.2$ and used two values of the
chemical potential of the left lead $\mu_L=0.4$ and $\mu_L=0.6$. For
$\mu_R=0.2$ and $\mu_L=0.4$ the bias window contains only the 1-st and
the 2-nd excited configurations with $N=1$, Fig.\ 5(b).  The ground
states for $N=1$ and $N=2$ are instead located below and above the bias
window, respectively. Consequently the steady state current is very small.
When $\mu_L$ increases to 0.6 the ground state configuration with $N=2$ enters the
bias window and the current increases, Fig.\ 6(a).

To analyze the transient regime we split the current into contributions
given by the ground state and excited states with 1 electron (see
Fig.\ 6(b)).  When $\mu_L=0.4$ the 1-st and 2-nd excited state carry
currents exceeding the current associated to the ground state, which
survive all the way to the steady state.  The current corresponding to
the 2-nd excited state is smaller than the current of the 1-st excited
state, but comparable to that of the ground state. This is explained
by the strength of the coupling coefficients shown in  Fig.\ 5(a), the
2-nd single-particle state being stronger coupled to the leads. The
remaining higher excited states give oscillating and fast decaying
transient currents.  In Fig.\ 6(c) $\mu_L=0.6$ and therefore higher
excited states enter the bias window; their transient currents are still
decaying but at a smaller rate. Comparing with Fig.\ 6(a) it in clear
that the transient regime is dominated by excited states.

\begin{figure}[tbhp!]
\includegraphics[width=0.45\textwidth]{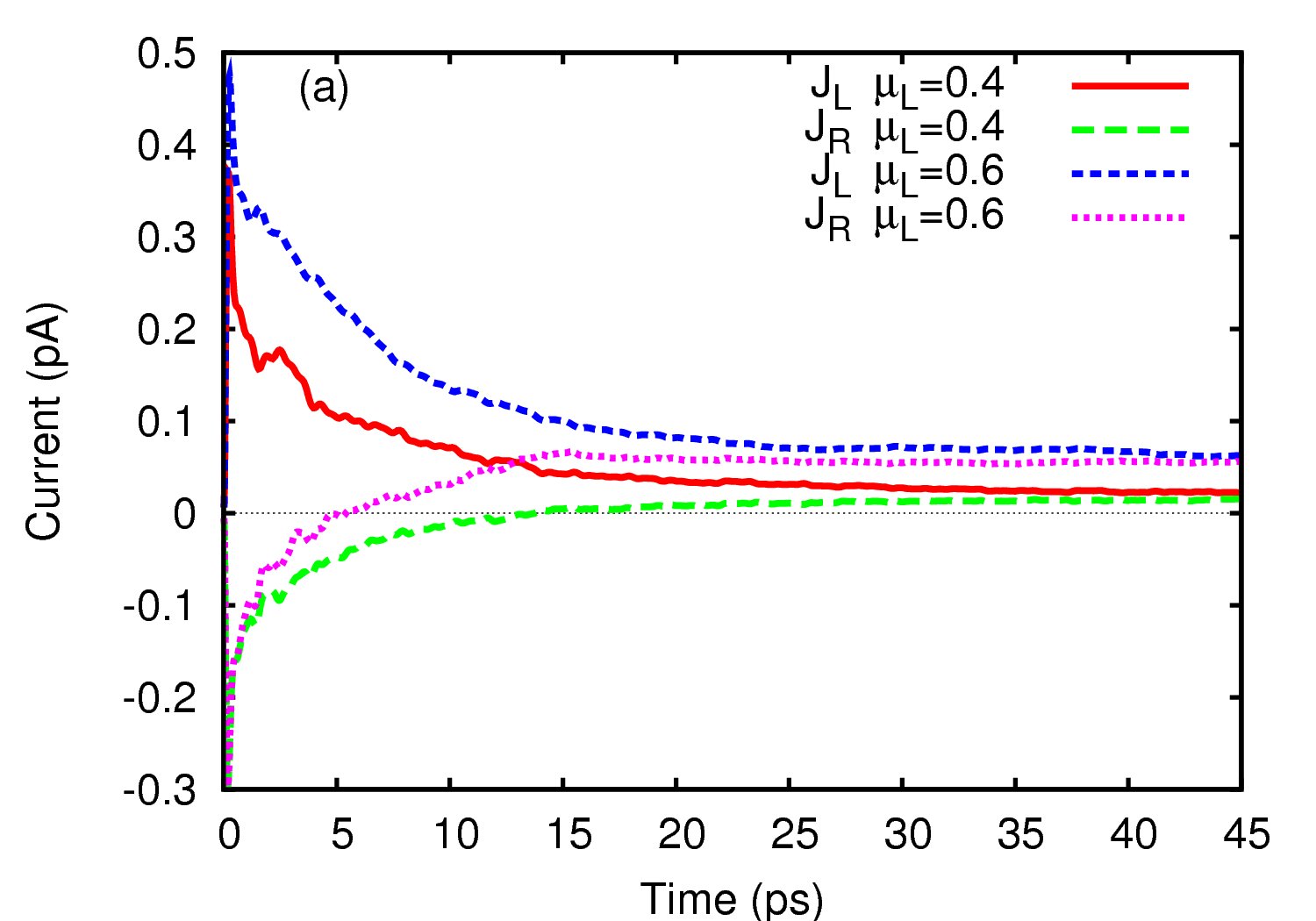}
\includegraphics[width=0.45\textwidth]{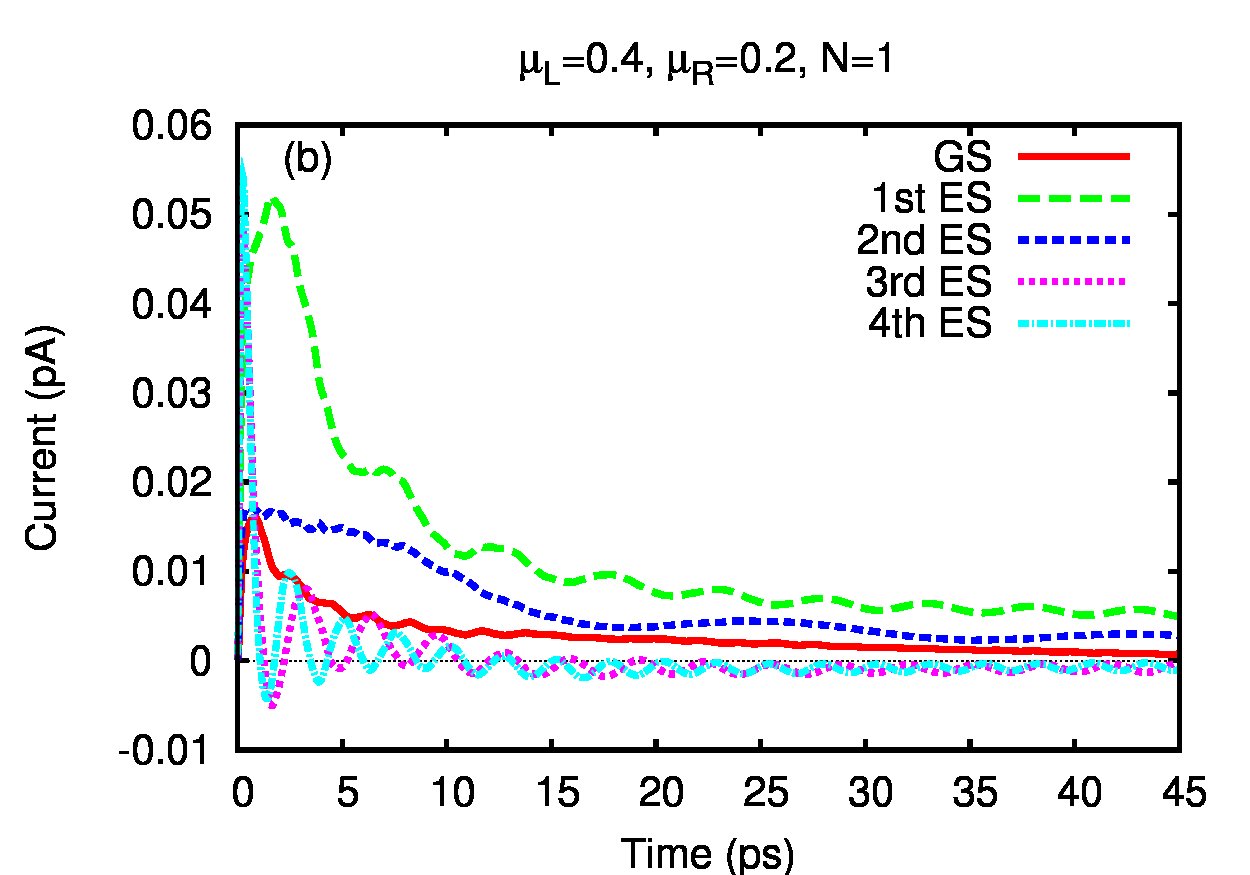}
\includegraphics[width=0.45\textwidth]{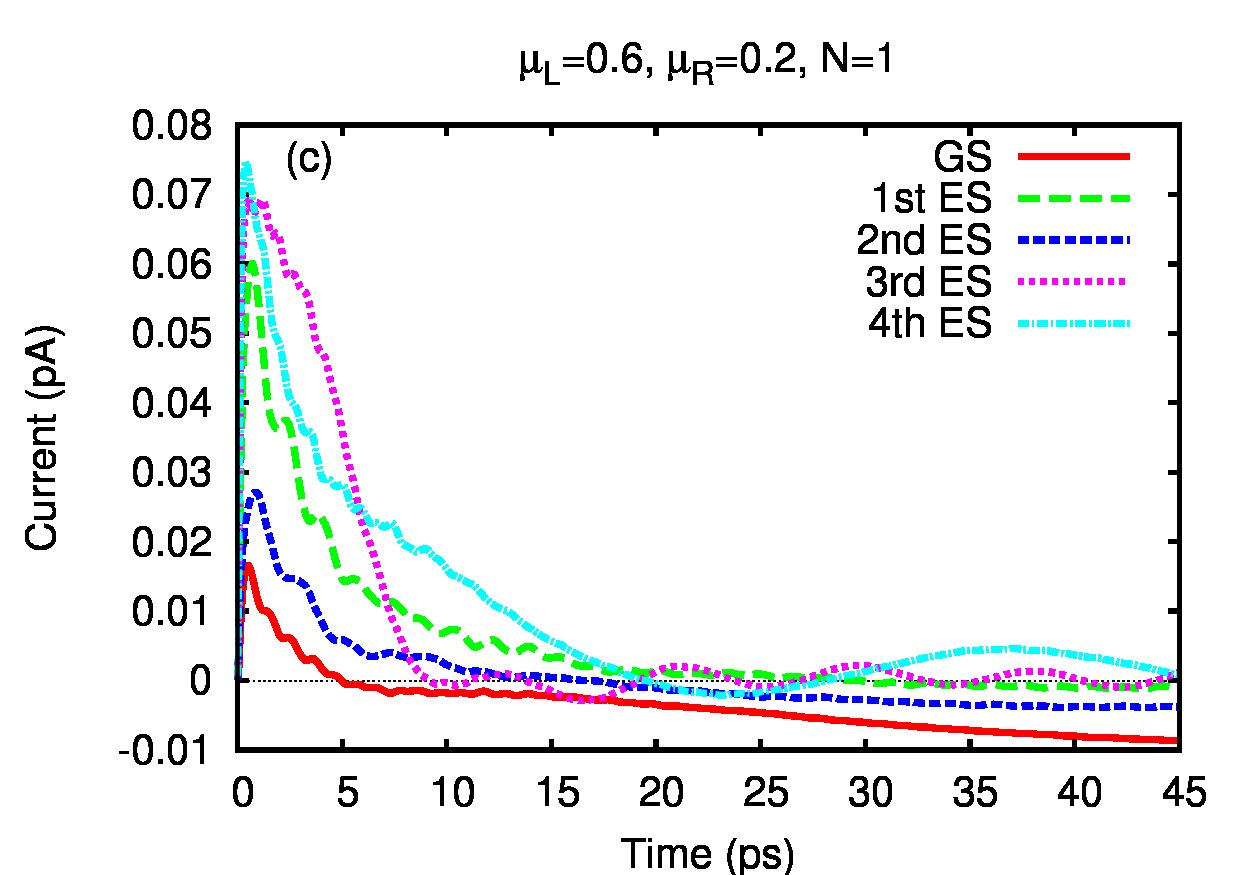}
\caption{ (a) The total currents in the left and right leads for $\mu_L=0.6$ and $\mu_L=0.4$,
while keeping $\mu_R=0.2$.
(b) The partial currents in the left lead for single-particle states when $\mu_L=0.4$ and $\mu_R=0.2$.
(c) The partial currents in the left lead for single-particle states when $\mu_L=0.6$ and $\mu_R=0.2$ }
\label{figure6}
\end{figure}

\begin{figure}[tbhp!]
\includegraphics[width=0.45\textwidth]{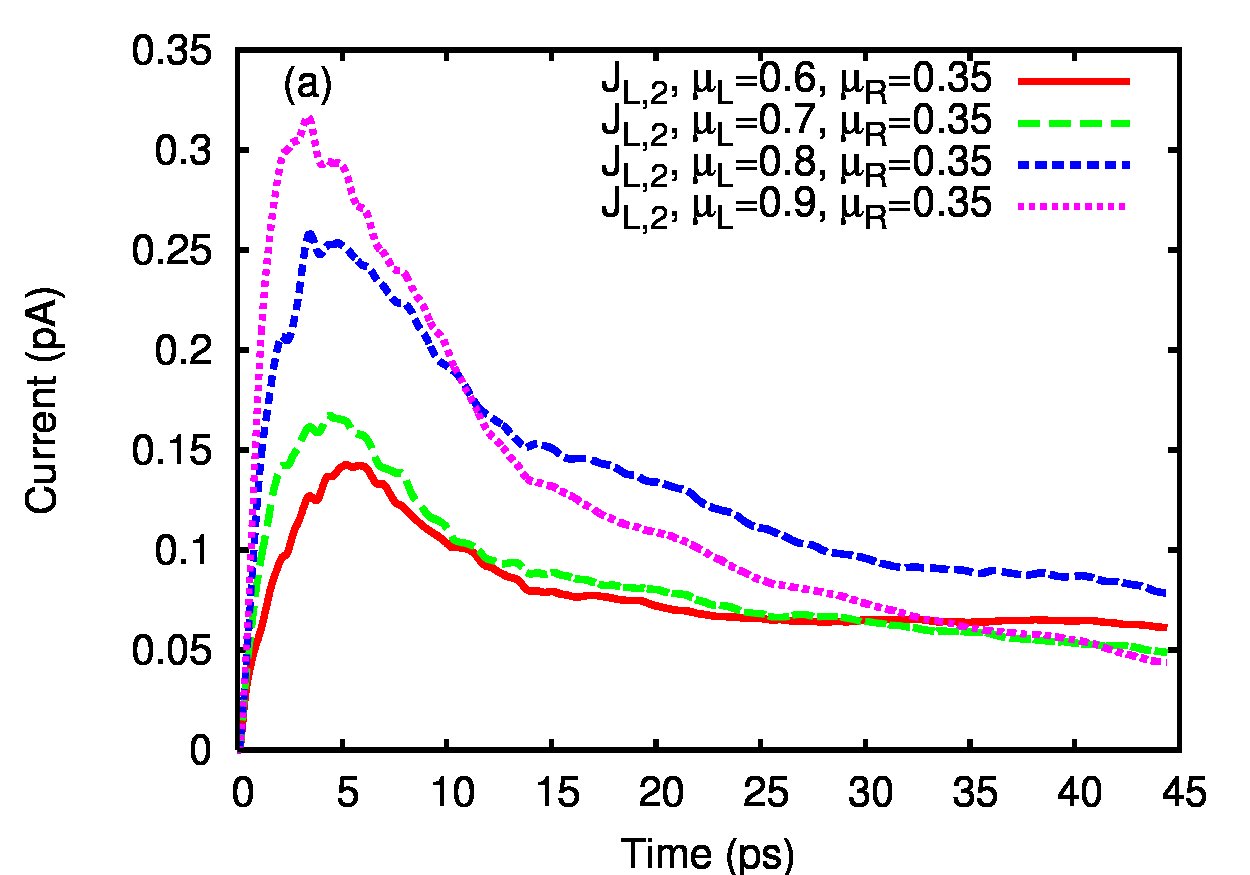}
\includegraphics[width=0.45\textwidth]{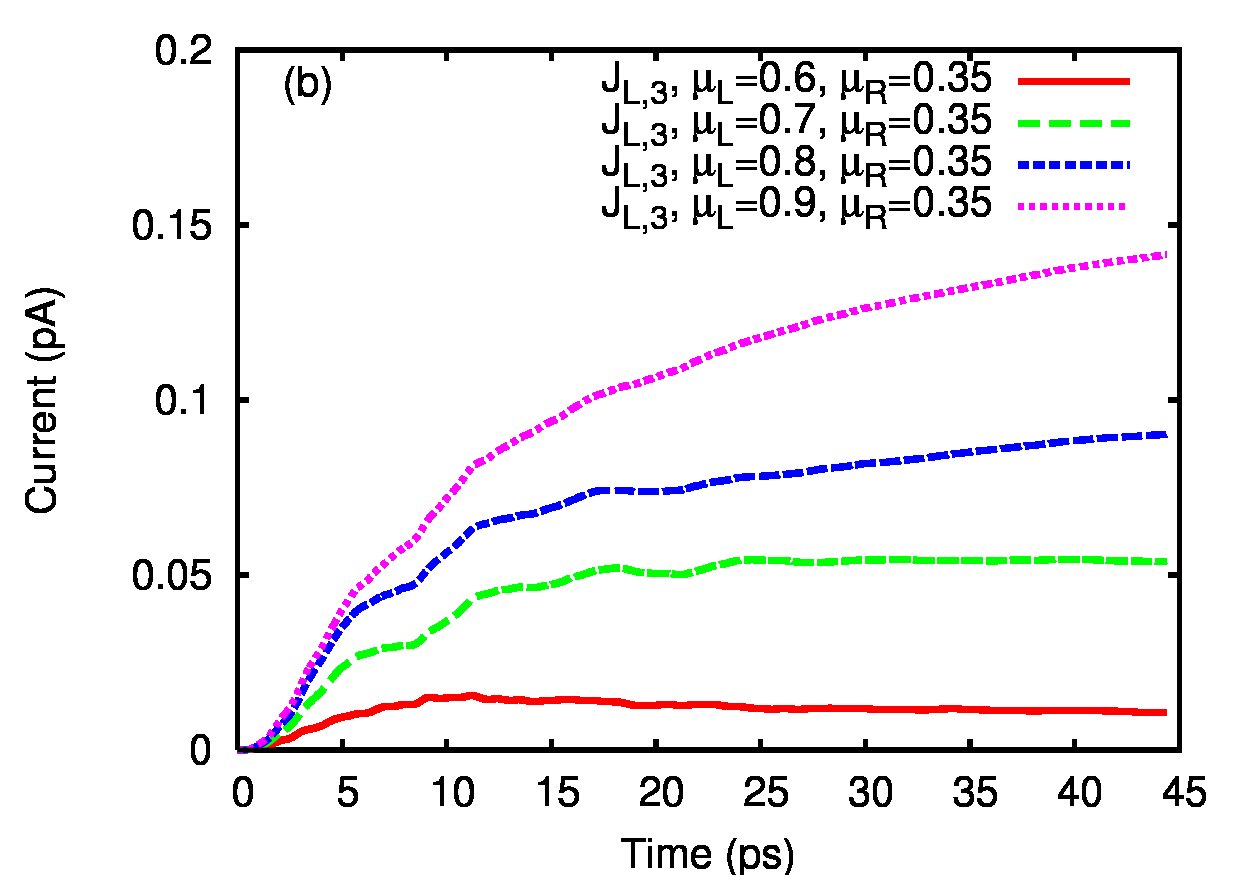}
\caption{(a) The total current in the left lead carried by all many-body
configurations with $N=2$, for increasing values of $\mu_L$ ({\it i.\,e.}
0.6,0.7,0.8 and 0.9) and $\mu_R=0.2$.  (b) The same for $N=3$. }
\label{figure7}
\end{figure}

Next we discuss currents associated with states having 2 and 3 electrons.
We keep now fixed $\mu_R=0.35$ and again increase $\mu_L$ starting
with 0.6.  Fig.\ 7(a) shows the total currents in the left lead for $N=2$
and $N=3$. As the bias increases the transient currents are enhanced,
but they become comparable as the system approaches the steady state. In
Fig.\ 7(b) the total current on three particle states shows a different
behavior: the steady states value increases drastically when $\mu_L$ moves
up. To explain this one can look again at the diagram of the chemical
potentials, Fig.\ 5(b). At $\mu_L=0.6$ the 3-particle configurations are
above the bias window and as such they contribute less to the current. In
contrast, as $\mu_L$ increases the ground state configuration with $N=3$
enters the bias window, the window is closer to the excited states,
and thus the total current increases. Actually, for $\mu_L=0.8$ and
$0.9$ the current for $N=3$ does not reach the steady state in the time
interval considered.

\begin{figure}[tbhp!]
\includegraphics[width=0.45\textwidth]{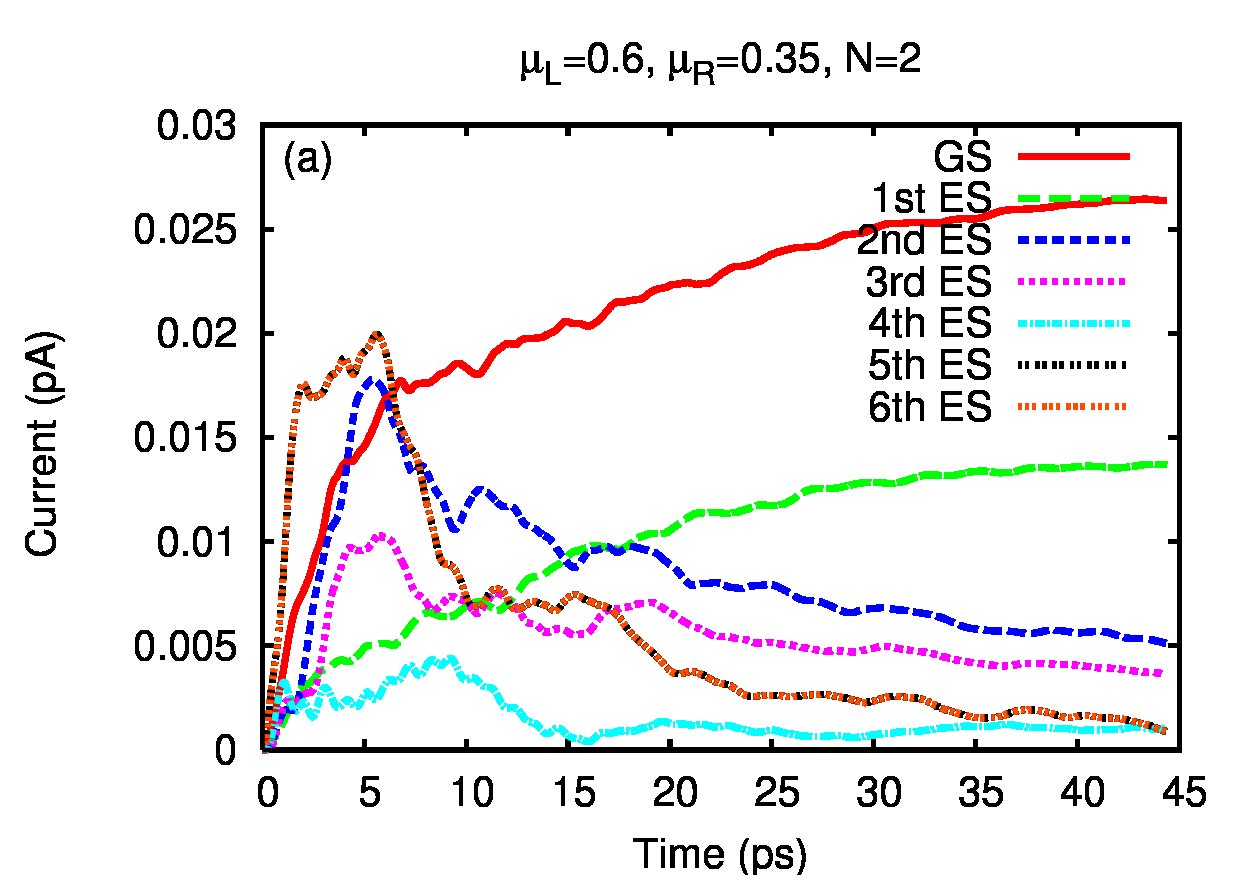}
\includegraphics[width=0.45\textwidth]{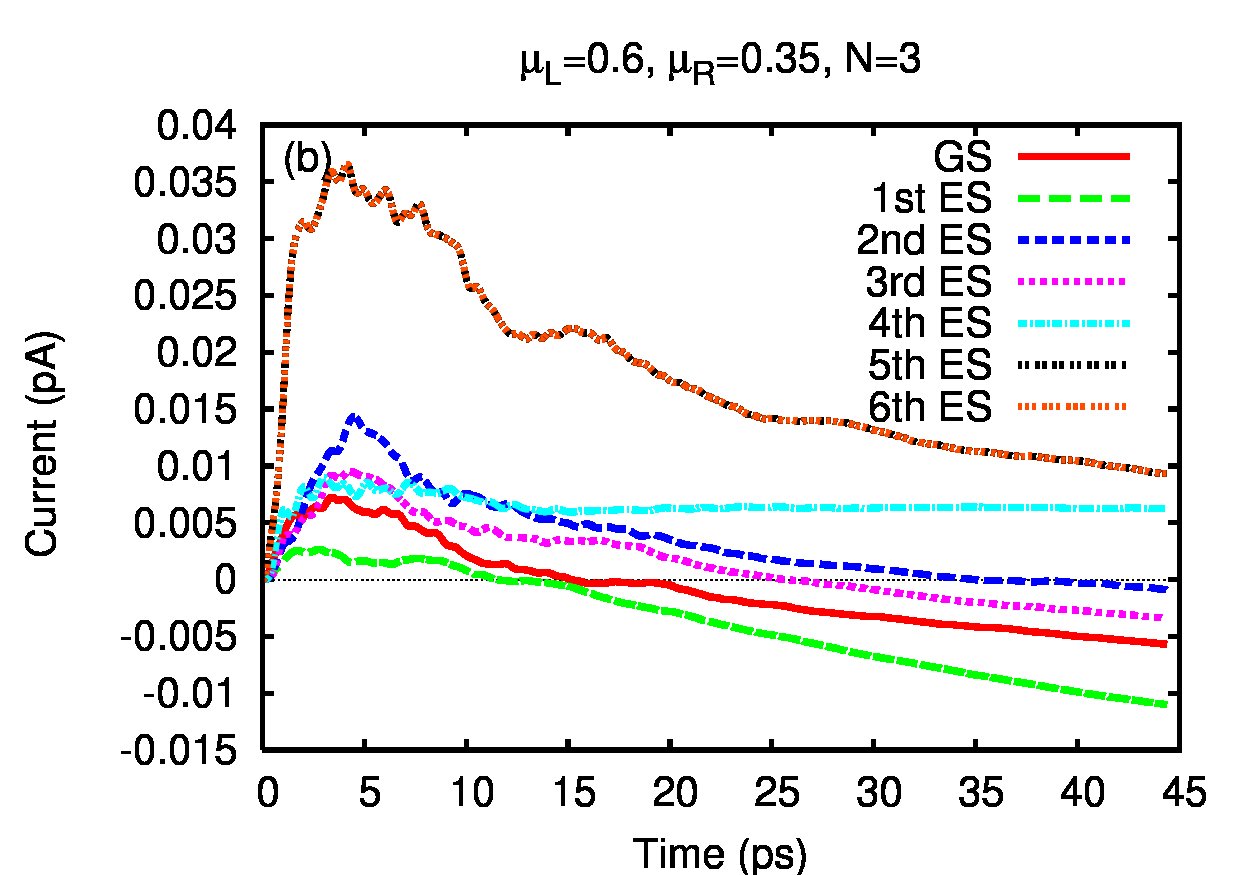}
\caption{(a) The total current in the left lead carried by all many-body
configurations with $N=2$ at $\mu_L=0.6$. (b) The same for  $\mu_L=0.9$.
Other parameters $\mu_R=0.35$. }
\label{figure8}
\end{figure}

Now we look at the contribution of the excited states with $N=2$ for two
cases, $\mu_L=0.6$ and $\mu_L=0.9$.  Again, the inspection of the diagram
in Fig.\ 5(b) predicts the results of Fig.\ 8. When $\mu_L=0.6$ there
is just one excited configuration within the bias window, in addition to
the ground state. In Fig.\ 8(a) we see that in the steady state these two
configurations give significant contributions to the current, whereas the
higher excited states play a role only in the transient regime. Fig.\ 8(b)
shows that at $\mu_L=0.9$ the currents of the excited states and of the
ground state are decreasing, some of them reaching even negative values
towards the steady state. This happens because the bias window includes
now the ground state with $N=3$ and the excited states with $N=2$
deplete in the favor of the ground state.

The sign of the current carried by states with $N$ particles depends
on the placement of the corresponding ground state chemical potential
relatively to the bias window. For example if we fix $\mu_L=1.5$
and $\mu_R=0.65$ we obtain $\mu_2^{(0)}<\mu_L$. Fig.\ 9(a) shows the
$N$-particle currents when the sample initially contains two electrons
in the ground state. This initial state evolves faster to the steady
state than the empty system.  While for $N=3$ the current in the left
lead is positive, for both $N=2$ and $N=1$ the currents are negative.
The charge residing on each $N$-particle state and the total charge are
shown in Fig.\ 9(b). Since single-particle configurations are unlikely
their occupation vanishes. The total charge accumulated on the $N=3$
states increases up to 2, while the total charge on the $N=2$ states
decreases from 2 to 0.75. The sign of the current for $N=2$ becomes
positive when $\mu_R$ is lowered to 0.2, Fig.\ 9(c), and exceeds the
current carried by the states with $N=3$.  This is because the 1-st and
the 2-nd SES practically determine the ground state with two electrons
and thus $\mu_2^{(0)}$, and also because the 1-st SES is strongly coupled
to the leads.  However, the current with $N=1$ is still negative.
\begin{figure}[tbhp!]
\includegraphics[width=0.45\textwidth]{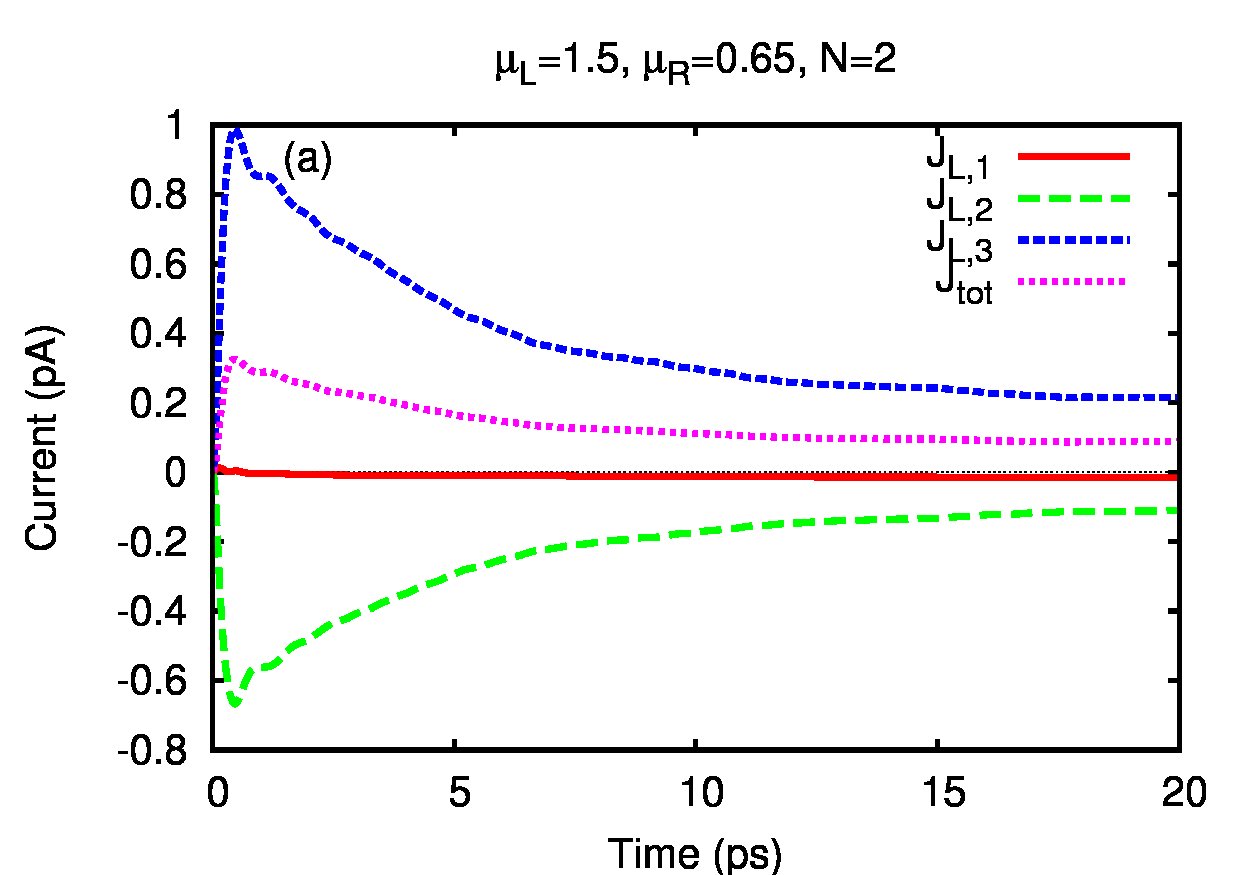}
\includegraphics[width=0.45\textwidth]{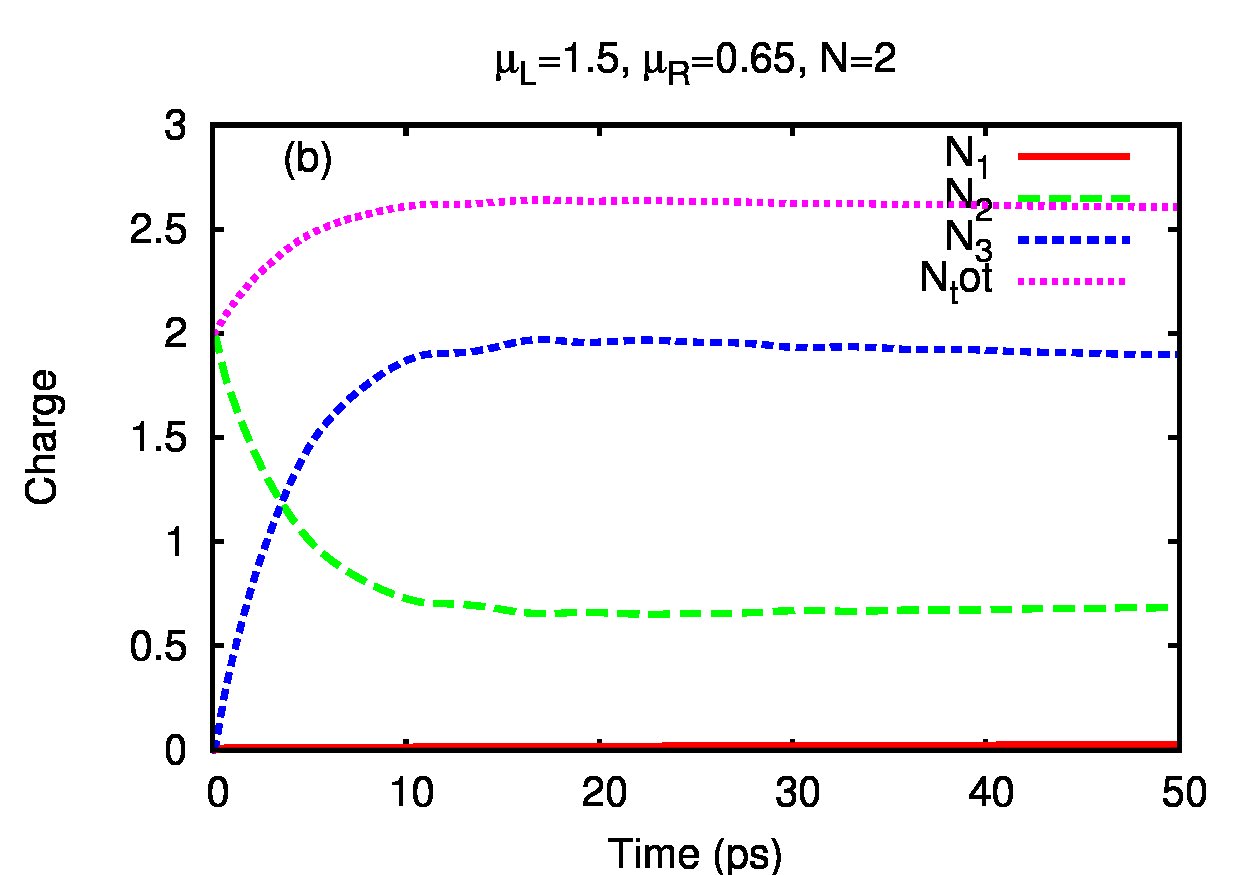}
\includegraphics[width=0.45\textwidth]{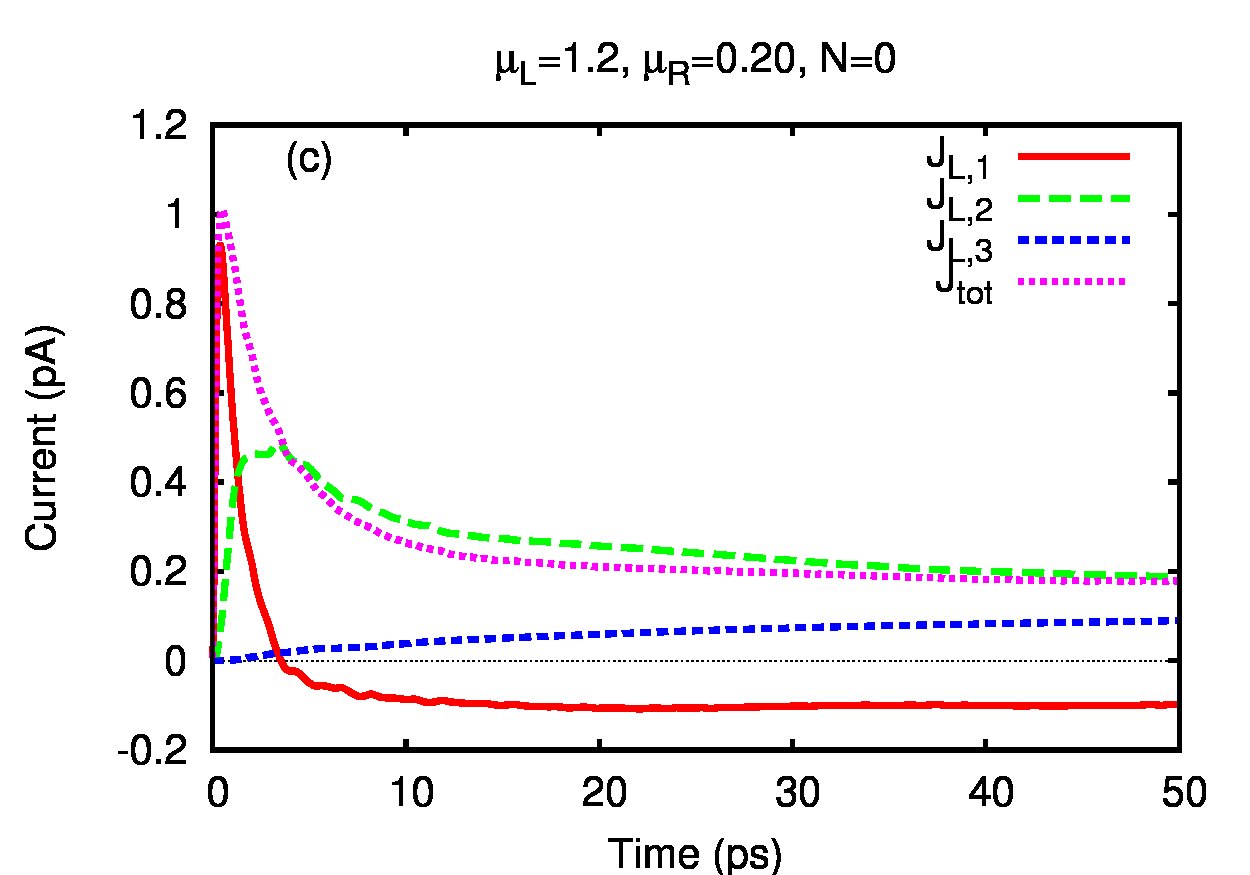}
\includegraphics[width=0.45\textwidth]{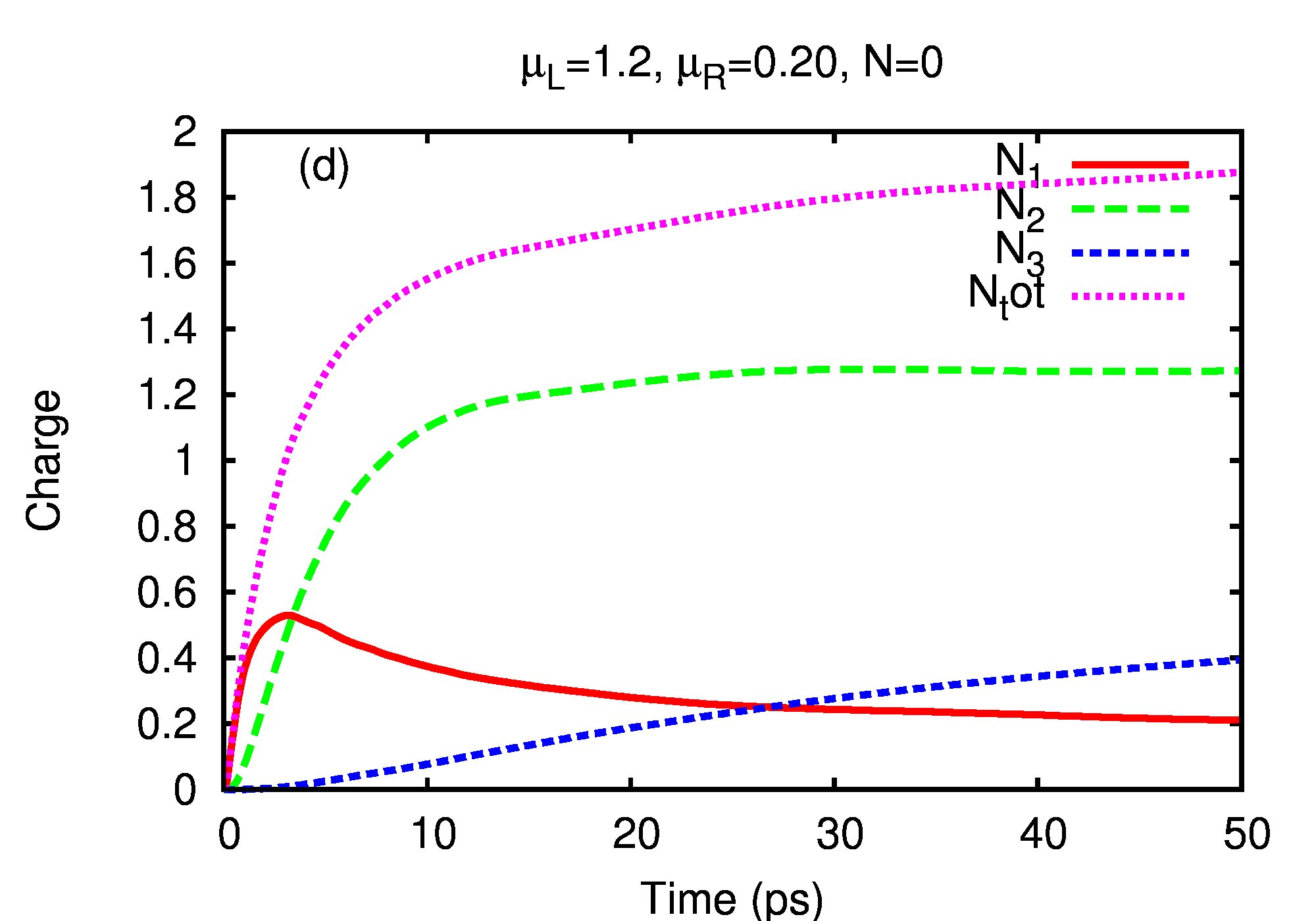}
\caption{(a) The total current in the left lead carried by $N$-particle
states and the total charge.  for $\mu_L=1.5$ and for $\mu_R=0.65$.
(b) The occupation number of the $N$-particle states.  (c) The total current
in the left lead carried by $N$-particle states for $\mu_L=1.2$ and $\mu_R=0.2$.
(d) The occupation number of the $N$-particle states and the total charge. }
\label{figure9}
\end{figure}

\subsection{Parabolic quantum wire}

In this subsection we apply the GME with Coulomb interaction to describe
the transport through a short quantum wire of length $L_x=300$ nm with a
parabolic confinement in the $y$-direction perpendicular to the direction
of transport.  The contact ends of the isolated wire at $\pm L_x/2$
are described by hard walls. This is now a continuous model, where a
large functional basis is used to expand the eigenfunctions of the system
in. In a similar manner we use a functional basis with complete truncated
sets of continuous and discrete functions to expand the eigenfunctions
of the semi-infinite parabolic leads in.  To show that we can describe
the combined geometrical effects imposed on the system by it's geometry
and an external perpendicular magnetic field we place the quantum wire
is in an external magnetic field of strength $1.0$ T. The characteristic
confinement energy is given by $\hbar\Omega_0=1.0$ meV. We assume GaAs
parameters with $m^*=0.067m_e$, $\kappa = 12.4$ meV. The magnetic length
modified by the parabolic confinement is $a_w=\sqrt{\hbar/(m^*\Omega_w)}$,
with $\Omega_w^2=\Omega_0^2+\omega_c^2$.  and the cyclotron frequency
$\omega_c=eB/(m^*c)$. At $B=1.0$ T, $a_w=23.87$ nm.  The semi-infinite
leads having the same parabolic confinement and being subject to the same
external perpendicular magnetic field have a continuous energy spectrum
with discrete Landau sub-bands.

The Coulomb potential in Eq.\ (\ref{Vcoul}) in the 2D wire is described by
\begin{equation}
      u(\mathbf{r}-\mathbf{r}')=
      \frac{e^2}{\kappa\sqrt{(x-x')^2+(y-y')^2+\eta^2}},
\end{equation}
with the small convergence parameter $(\eta /a_w) = 0.01$ to facilitate the
two-dimensional numerical integration needed for the matrix elements (\ref{Vcoul}).

After the GME (\ref{GMEfin}) has been transformed to the interacting
many-electron basis by the unitary transformation obtained by
the diagonalization of $H_S$ (\ref{Hsample}) we truncate the RDO
(\ref{redrho}) to 32 MES. For the bias range $0.0 \leq \Delta\mu =
\mu_L-\mu_R \leq 1.7$ meV used here 10 SES are sufficient to obtain these
lowest 32 states with good accuracy. We will be omitting singly occupied
states of high energy that should not be relavant for the parameters here.
The natural strength of the Coulomb interaction will only give us MES
that are occupied by one or two electrons in the energy range 0 to 6
meV covered by the 32 MES.

Since in the partitioning approach $[H_S,H_L]=0$ we have to construct $T^l_{qn}$ as
a {\it non-local} overlap of $\phi_n$ and $\psi_q^{L,R}$ on the contact
regions ${\cal C}_{l}, \, l=L,R$:\,\cite{NJP2}

\begin{equation}\label{Tqn_c}
T^l_{qn}=\int_{{\cal C }_l}d{\bf r}d{\bf r'}
\left( \psi^{*}_{ql}({\bf r})\phi_n({\bf r}) g^l_{qn}({\bf r},{\bf r'}) +h.c. \right).
\end{equation}
\begin{align}
      g^l_{qn} ({\bf r},{\bf r'}) =
                   g_0^l&\exp{\left[-\delta_1^l(x-x')^2-\delta_2^l(y-y')^2\right]}\nonumber\\
                   \times &\exp{\left(\frac{-|E_n-\varepsilon_{ql}|}{\Delta_E^l}\right)} \,.
\label{gl}
\end{align}
As before $\varepsilon_{ql}$ is the energy spectrum of lead $l$, and
$E_n$ is the energy of the SES numbered by $n$ in the quantum wire.
The quantum number $q$ for the states in leads represents both the
discrete Landau band number and a continuous quantum number that can
be related to the momentum of a particular state.  Here we use the
parameters $\delta_1a_w^2=\delta_2a_w^2=0.25$, $\Delta_E^{LR}=0.25$ meV,
and $g_0^{LR}=40$ meV for $B=1.0$ T. The domain of the overlap integral
for the leads is $\pm 2a_w$ into the lead or the system for $x$ and $x'$
from each end of the wire at $\pm L_x/2$ and between $\pm 4a_w$ for $y$
and $y'$, see Ref.\ (\onlinecite{NJP2}) for an exact definition.  All the
SES will be coupled to the leads, but the coupling strength will depend on
the character of the SES, whether it is an edge- or bulk state and other
finer geometrical details that is brought about by the magnetic field.

\begin{figure}[htbq]
      \includegraphics[width=0.48\textwidth]{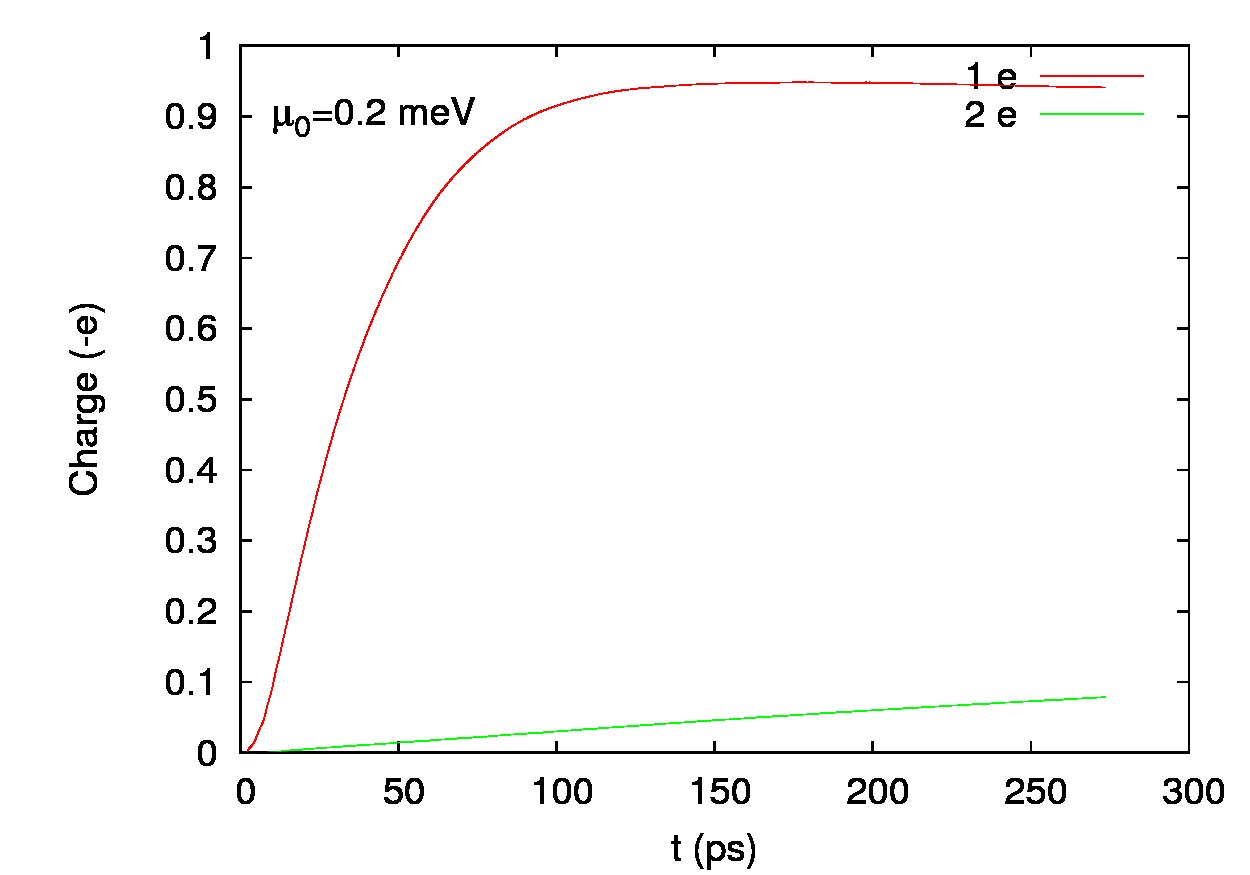}
      \includegraphics[width=0.48\textwidth]{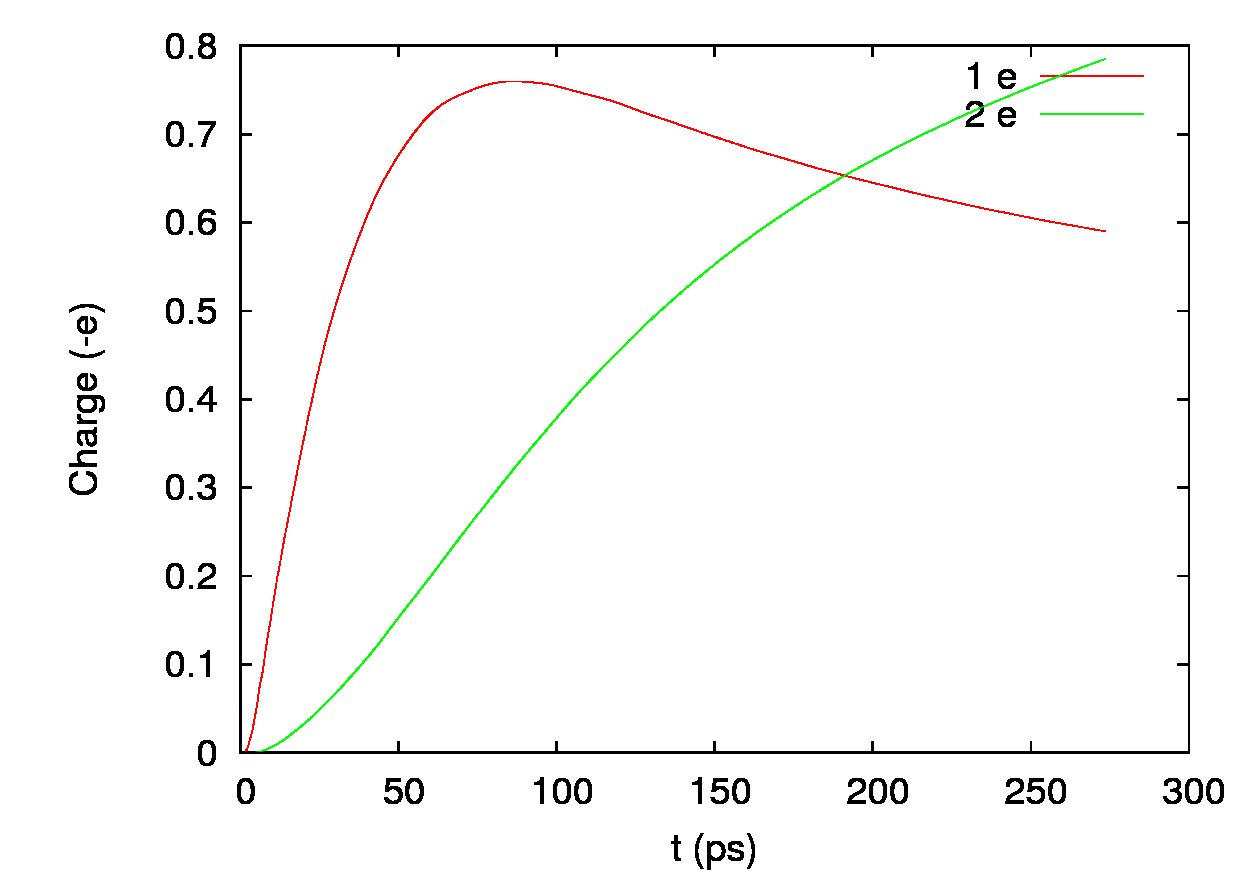}
      \caption{(Color online) The total charge residing in one- and two-electron
               states as a function of time for two different values of the bias
               $\Delta\mu$. $B=1.0$ T, $L_x=300$ nm, $\hbar\Omega_0=1.0$ meV.}
      \label{1e2evg}
\end{figure}

The right chemical potential $\mu_R$ is held at $1.4$ meV and the transport
properties are calculated for different values of the bias $\Delta\mu$ by
varying $\mu_L$. Figure \ref{1e2evg} compares the total occupation of all one-
electron and two-electron MES for the interacting system at two different values
of the bias.
At, $\Delta\mu =0.2$ meV we see that almost solely one-electron states are occupied,
while for $\Delta\mu =1.2$ meV initially it is likely to have one-electron states
occupied, but very soon the occupation of the two-electron states becomes as
probable with the likelihood of the occupation of the one-electron states fast
reducing with time. We also have to admit here that even though the steady state
value of the total current through the system can be deduced by the values of
the current at 270 ps, the charging of the system takes much longer time, since
we are using here a very weak coupling to the leads that mimics a tunneling regime.

\begin{figure}[htbq]
      \begin{center}
      \includegraphics*[width=0.48\textwidth,angle=0]{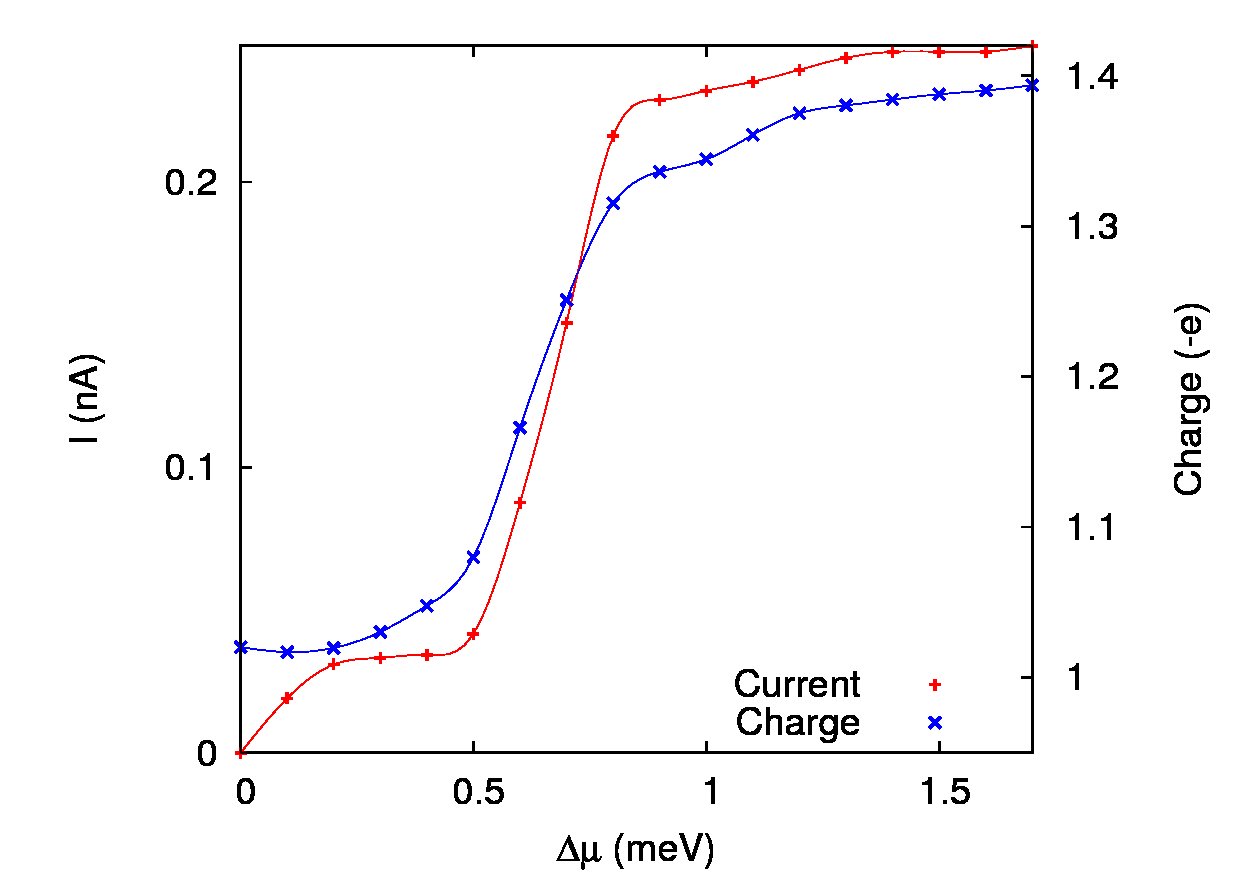}
      \end{center}
      \caption{The total steady state current for interacting 10 SES, and the
               total charge at $t=270$ ps.
               for different values of the bias $\Delta\mu$.
               $B=1.0$ T, $L_x=300$ nm, $\hbar\Omega_0=1.0$ meV.}
\label{0Imu10ses}
\end{figure}

If we now use the average value of the current in the left and right leads
at $t=270$ ps as a measure of the steady
state current we get the information displayed in Fig.\ \ref{0Imu10ses},
where the steady state value of the current is shown for the interacting
system as a function of the bias and compared to the charge in the system.
We have a clear Coulomb blocking in the interacting system.  In the
case of a non-interacting system the lack of a gap between the one- and
two electron MES and a strong mixing of the energy regimes of two- and
three-electron states the two-electron plateau only appears as a small
shoulder. The 32 MES selected here include no three-electron or MES with
higher number of electrons. It should be mentioned here that a different
choice of the right bias $\mu_R$ can result in the system charging faster
and thus at the same time the total current through it being smaller.
This comes from the fact that the states have a different coupling to
the leads and the time range shown here is very much in the transient-
or it's long exponential decay regime.

\begin{figure}[htbq]
      \begin{center}
      \includegraphics*[width=0.48\textwidth,angle=0]{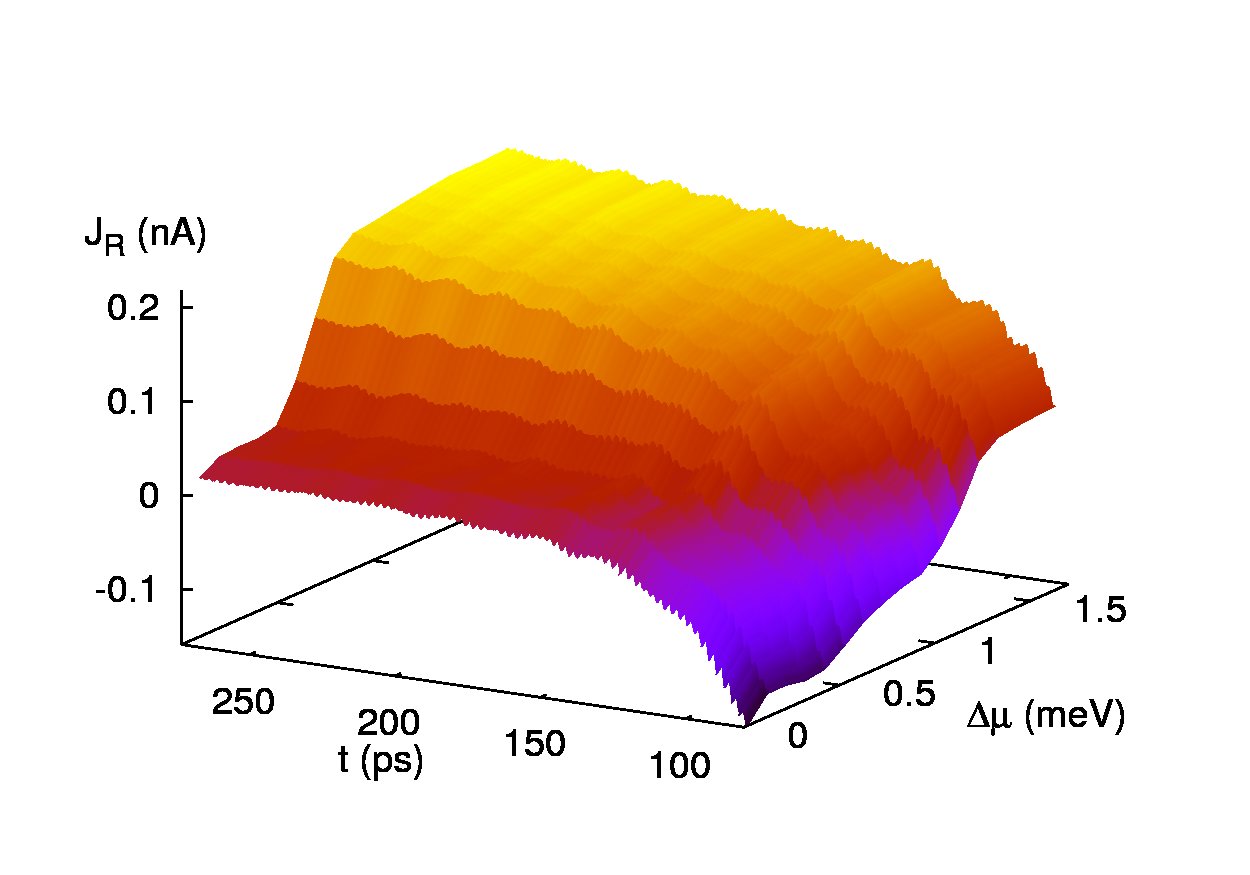}
      \end{center}
       \caption{The total current in the right lead for interacting and non-interacting 10 SES
               as a function of the bias $\Delta\mu$ and time.
               $B=1.0$ T, $L_x=300$ nm, $\hbar\Omega_0=1.0$ meV.}
\label{3DJR}
\end{figure}

Figure \ref{3DJR} displaying the current in the right lead gives an idea
how the Coulomb blocking plateau appears after the transition regime.
The transition regime where the right current goes negative, {\it i.\,e.}\ where it supplies
charge to the system is partially truncated from the figure.

\section{Summary and conclusions}

We calculated time-dependent currents in open mesoscopic systems
composed by a sample attached to two semi-infinite leads, by solving
the generalized master equation for the reduced density operator acting
in the Fock space of the sample.  This is the natural framework for
including the Coulomb electron-electron interaction in the sample,
which is the main achievement of this work. The Coulomb interaction
is treated in the spirit of the exact diagonalization method, {\it
i.\,e.}\ in a pure many-body manner.  The interacting many-body states
of the sample are expanded in the basis of non-interacting ``bit-string''
states with unspecified number of electrons. We believe our method is
a viable alternative to a recent approach based on a time-dependent
density-functional model.\cite{Kurth,Myohanen,Kurth2} We used three
sample models, a short 1D wire with 5 sites, but also a larger 2D lattice
with 120 sites and a continuous model, whereas the cited group used much
smaller samples even with no structure.\cite{Kurth2}

Indeed, due to computational limitations we could use only a restricted,
effective number of many-body states in the GME, between 30-50 depending
on the model, from the bottom of the energy spectrum.  We chose the bias
window $[\mu_R,\mu_L]$ and the strength of the sample-leads coupling
parameters $V_{R,L}$ such that only the effective states contribute
to the transport of electrons through the sample, whereas the states
with higher energy are unreachable by the electrons.  Consequently the
number of electrons in the sample can be only up to 3 or 4.

We could calculate the contribution to the charge and currents in the
sample and in the leads respectively, corresponding to any particular
many-body state. We use the 1D chain as a toy model to emphasize the
dominant role of the excited states in the transient regime and the
onset of the Coulomb blockade in the steady state. A similar 1D model
with 4 sites 1D has been considered recently by My\"oh\"anen {\it et
al.}\cite{Myohanen}

As shown also in our previous works on time-dependent transport in
non-interacting systems the GME method includes information on the energy
structure of the sample, but also on the geometrical properties reflected in
the wave functions and sample-lead contacts.\cite{NJP1,NJP2,Valim2}
Here we illustrate these aspects, in the interacting case, for two
nanosystems: a two-dimensional quantum dot described by a lattice
Hamiltonian and a short parabolic quantum wire. The time-dependent
occupation of specific many-body states was thoroughly analyzed, for
different values of the chemical potentials of the leads.  It turned out that
the excited states with $N$ electrons contribute to the steady state
currents if the ground state configuration with $N+1$ particles is not
available for transport. However, if $\mu^{(0)}_N<\mu_R$ and at the
same time $\mu^{(0)}_{N+1}$ lies within the bias window the excited states
with $N$ particles are active only in the transient regime and become
de-populated in the steady state regime. This behavior is of interest
in the excited-state spectroscopy experiments.\cite{Tarucha} To our
knowledge the time-dependent currents associated to excited states have
not been discussed theoretically so far.

\begin{acknowledgments}
The authors acknowledge financial support from the Development Fund of
the Reykjavik University Grant No.\ T09001, the Research and Instruments
Funds of the Icelandic State, the Research Fund of the University of
Iceland, the Icelandic Science and Technology Research Programme for
Postgenomic Biomedicine, Nanoscience and Nanotechnology, the National
Science Council of Taiwan under contract No.\ NSC97-2112-M-239-003-MY3.
V.M.\ also acknowledges the hospitality of the Reykjavik University,
Science Institute and the partial financial support from PNCDI2 program
(grant No.\ 515/2009) and grant No.\ 45N/2009.
\end{acknowledgments}
%

%
%

\bibliographystyle{apsrev}

\begin{thebibliography}{9}

\bibitem{Stefanucci}
G. Stefanucci, S. Kurth, A. Rubio and E. K. U. Gross, Phys. Rev. B 77,
075339 (2008).

\bibitem{Moldo}
V. Moldoveanu, A. Manolescu and V. Gudmundsson, Phys. Rev. B {\bf 76},
085330 (2007)

\bibitem{Xiao}
X. Zheng, F. Wang, C. Y. Yam, Y. Mo, and G.H. Chen, Phys. Rev. B {\bf 75},
195127 (2007).

\bibitem{Gudmundsson}
V. Gudmundsson, G. Thorgilsson, C-S Tang, and V. Moldoveanu, Phys. Rev. B
77, 035329 (2008).

\bibitem{Harbola}
U. Harbola, M. Esposito, and S. Mukamel, Phys. Rev. B {\bf 74}, 235309
(2006).

\bibitem{Welack1}
S. Welack, M. Schreiber, and U. Kleinekath\"{o}fer, J. Chem. Phys. 124,
044712 (2006)

\bibitem{NJP1}
V. Moldoveanu, A. Manolescu and V. Gudmundsson,  New J. Phys. 11, 073019
(2009).

\bibitem{NJP2}
V. Gudmundsson, C. Gainar, C-S Tang, V. Moldoveanu, A. Manolescu, to
appear in  New J. Phys.

\bibitem{Kurth}
S. Kurth, G. Stefanucci, C.-O. Almbladh, A. Rubio, and E. K. U. Gross,
Phys. Rev. B {\bf 72}, 035308 (2005).

\bibitem{Myohanen}
P. My\"{o}h\"{a}nen, A. Stan, G. Stefanucci, and R. van Leeuwen, Phys. Rev. B
{\bf 80}, 115107 (2009), Europhys. Lett. {\bf 84}, 67001 (2008).

\bibitem{Tarucha}
T. Fujisawa, D. G. Austing, Y. Tokura, Y. Hirayama, and S. Tarucha,
J. Phys.: Condens. Matter {\bf 15}, R1395 (2003).

\bibitem{Langreth}
D. C. Langreth and P. Nordlander, Phys. Rev. B {\bf 43}, 2541 (1991)

\bibitem{Gurvitz}
S. A. Gurvitz and Ya. S. Prager, Phys. Rev. B {\bf 53}, 15932 (1996).

\bibitem{real}
J. K\"{o}nig, H. Schoeller, and G. Sch\"{o}n, Phys. Rev. Lett. {\bf 76},
1715 (1996); J. K\"{o}nig, J. Schmid, H. Schoeller, and G. Sch\"{o}n,
Phys. Rev. B {\bf 54}, 16 820 (1996).

\bibitem{Becker}
D. Becker and D. Pfannkuche, Physical Review B {\bf 77} 205307 (2008).

\bibitem{Nakajima}
S. Nakajima, Prog. Theor. Phys. {\bf 20}, 948 (1958).

\bibitem{Zwanzig}
R. Zwanzig, J. Chem. Phys. {\bf 33}, 1338 (1960).

\bibitem{Timm}
C. Timm, Phys. Rev. B {\bf 77}, 195416 (2008).

\bibitem{Li}
X. Q. Li and Y. J. Yan, Phys. Rev. B {\bf 75}, 075114 (2007).

\bibitem{Esposito}
M.Esposito and M. Galperin, Phys. Rev. B {\bf 79}, 205303 (2009).

\bibitem{Darau}
D. Darau, G. Begemann, A. Donarini, and M. Grifoni, Physical Review B
{\bf 79} 235404 (2009).




\bibitem{Caroli}
C. Caroli, R. Combescot, P. Nozieres, and D. Saint-James, J. Phys. C {\bf 4}, 916 (1971).

\bibitem{Vaz}
E. Vaz, J. Kyriakidis, J. Chem. Phys. {\bf 129}, 024703 (2008)

\bibitem{Kouwenhowen}
L.P. Kouwenhouven et al., in Mesoscopic Electron Transport, edited by
L.L. Sohn, L.P. Kouwenhouven, and G. Schön, NATO Advanced Study Institute,
Series E, Vol. 345 (Kluwer, Dordrecht, 1997).

\bibitem{Valim2}
V. Moldoveanu, A. Manolescu and V. Gudmundsson,  Phys. Rev. B {\bf 80}, 205325 (2009).

\bibitem{Kurth2}
S. Kurth, G. Stefanucci, E. Khosravi, C. Verdozzi, ad E. K. U. Gross, e-print arXiv:0911.3870
[cond-mat.mes-hall] (2009).

\end{thebibliography}

\end{document}